\documentclass[aps,prb,twocolumn,nofootinbib,superscriptaddress]{revtex4-2}
\usepackage{amssymb,amsmath}
\usepackage{graphicx}
\usepackage{xcolor}
\usepackage{hyperref}
\usepackage{listings}

\newcommand{\bra}[1]{\langle #1|}
\newcommand{\ket}[1]{|#1\rangle}
\begin{document}

\title{Cross-Resonant Gates in Hybrid Fluxonium-Transmon Systems}
\author{Nikola D. Dimitrov}
\email{ndimitrov@wisc.edu}
\affiliation{Department of Physics, University of Wisconsin-Madison, Madison, WI, USA}
\affiliation{SEEQC, Inc., Elmsford, NY, USA}
\author{Chen Wang}
\affiliation{Department of Physics, University of Massachusetts-Amherst, Amherst, MA, USA}
\author{Vladimir E. Manucharyan}
\affiliation{Institute of Physics, Ecole Polytechnique Federale de Lausanne, Lausanne, Switzerland}
\author{Maxim G. Vavilov}
\affiliation{Department of Physics, University of Wisconsin-Madison, Madison, WI, USA}
\date{\today}
\begin{abstract}
We propose a scalable dual-species fluxonium-transmon system that utilizes a central transmon to mediate high-fidelity gates and parity checks between two fluxonium qubits without the need for strong non-local interactions. This approach suppresses unwanted long-range interactions, which is critical for developing larger quantum processors. First, we analyze the performance of cross-resonance (CR) \textsc{cnot} gates between a fluxonium and a transmon. We show that even in the presence of a spectator qubit, these gates maintain high fidelity with coherent errors on the order of $10^{-5}$.
We then demonstrate that these gates, when applied sequentially, enable high-fidelity parity checks and logical fluxonium-fluxonium \textsc{cnot} gates. In addition, the central transmon can facilitate the readout of the neighboring fluxoniums, consolidating multiple critical functions into a single ancilla. Our work establishes the viability of a dual-species architecture as a promising path toward fault-tolerant quantum computation.
\end{abstract}
\maketitle

\section{Introduction}
Utility-scale quantum computing promises algorithmic speedups with many useful applications~\cite{shor1999polynomial, grover1996fast, aspuru2005simulated}. However, a major obstacle is that current quantum hardware has error rates per gate that are far too high for known algorithms to be effective~\cite{preskill2018quantum}. Quantum Error Correction (QEC) is considered a promising path to achieving sufficiently low logical error rates necessary for real-world applications~\cite{gottesman2009introduction}. QEC codes rely on long-coherence qubits and high-fidelity gates to overcome the intrinsic physical error thresholds and enable favorable scaling of logical error rates. Many proposed QEC codes, including those in the Calderbank--Shor--Steane class~\cite{calderbank1996good, calderbank1997quantum, steane1996simple}, require measuring combined qubit parities to determine syndromes that indicate errors. As such, achieving higher-fidelity two-qubit gates and pairwise parity checks is crucial for realizing quantum algorithms with greater circuit depth.

Superconducting qubits have become one of the leading platforms for building large-scale quantum computers~\cite{arute2019supremacy, google2025willow, gao2025establishing}, and pushing the boundaries of QEC~\cite{Acharya2023, Krinner2022, Marques2022, zhao2022realization}. Much of this progress has been driven by the development of circuit quantum electrodynamics over the past decade~\cite{blais2004cavity, blais2007quantum, zhu2013circuit, krantz2019quantum, blais2021circuit}. Many ideas for maximizing the capabilities of superconducting qubits have been recently proposed and experimentally realized~\cite{brooks2013protected, casparis2018superconducting, earnest2018realization, brosco2024superconducting, joshi2021quantum, teoh2023dual, hays2025non, pechenezhskiy2020superconducting, mencia2024integer, wang2026proposal}.
In particular, two widely studied types of superconducting qubits are (\textit{i}) the transmon~\cite{koch2007charge}, known for its simplicity and reliability, and (\textit{ii}) the fluxonium~\cite{manucharyan2009fluxonium}, which benefits from higher coherence and anharmonicity. Recently, hybrid systems that combine both transmon and fluxonium qubits have shown promise for QEC applications~\cite{ciani2022microwave, heunisch2025scalable} by achieving high-fidelity gate operations~\cite{simakov2021high, ding2023high, singh2025fast}. Since the hardware requirements for QEC are extremely sensitive to the error rate of two-qubit operations~\cite{Sheldon2016, Jurcevic_2021, Kim2023}, the search for better systems and high-fidelity multiqubit operations remains one of the most critical steps toward realizing fault-tolerant quantum computation.

In this paper, we present a method for creating fast, high-fidelity entangling gates in dual-species fluxonium-transmon systems by utilizing the cross-resonance (CR) effect~\cite{Rigetti2010, Chow2011simple, magesan2020effective}. This technique avoids the need for dedicated physical couplers between qubits. Building on recent work that used selective darkening~\cite{de2010selective, deGroot2012SD} to achieve high-fidelity CR gates with two fluxoniums~\cite{nesterov2022cnot,dogan2023two, lin2025verifying,lin202524}, we show that this method is uniquely suited for the system consisting of a fluxonium qubit and a transmon qubit (FT). The higher frequency of the transmon qubit naturally provides a larger detuning with its fluxonium counterpart. This detuning allows us to achieve the conditions for selective darkening with an optimal mismatch of drive amplitudes. The transmon, which is the target qubit in a CR \textsc{cnot} gate, is subject to a weak drive, and is thus protected from leakage (unwanted excitation to higher energy states). The fluxonium, which acts as the control qubit, receives a strong drive at a frequency far detuned from its own. Because of this detuning, the drive has a minimal effect, ensuring that the control qubit remains unaffected by the \textsc{cnot} gate operation. Also, leakage errors from the stronger fluxonium drive are suppressed by the fluxonium's high anharmonicity. We simulate these CR \textsc{cnot} gates with coherent error well below $10^{-4}$ for gate times under 50 ns, a promising result for future superconducting quantum computing hardware.

Compared to the cross-resonance interaction between two qubits of the same species (e.g., two fluxoniums), our system's large detuning between the control fluxonium and the target transmon makes the gate optimization more complex. This sizable detuning results in a weak hybridization between fluxonium and transmon states, which, in turn, may require a stronger drive to implement fast entangling gates. While increasing the fluxonium-transmon coupling can strengthen this hybridization and allow for faster gates, it comes at the cost of a higher $ZZ$ interaction, a static qubit-qubit interaction that causes errors. Although these $ZZ$-effects can be fully corrected in a closed two-qubit system, their presence becomes a significant issue in a multi-qubit system with spectator qubits in arbitrary states, where perfect corrections are not possible. As a result, finding the proper trade-off between coupling strength, gate speed, and drive amplitude is a crucial step in ensuring the scalability of our interleaved fluxonium-transmon architecture. This interplay is thoroughly examined to find a performance sweet spot in Sec.~\ref{sec:III}.

To further investigate the scalability of our dual-species approach for QEC, we also consider the fluxonium-transmon-fluxonium (FTF) system~\cite{ding2023high, ciani2022microwave}. 
This architecture leverages the transmon's better-understood and more robust readout capabilities~\cite{shillito2022dynamics, cohen2023reminiscence, chen2023transmon, swiadek2024enhancing, kurilovich2025high, spring2025fast} to perform parity measurements and readout of neighboring fluxoniums. Additionally, the transmon can serve as an auxiliary qubit to facilitate logical operations between next-nearest-neighbor fluxonium pairs. All of these operations are feasible in a regime with transmon-fluxonium coupling set much stronger than residual fluxonium-fluxonium coupling. The benefit of this approach is the suppression of unwanted $ZZ$-type couplings arising from long-range qubit interactions. This reduced long-range interaction allows us to investigate gate performance in a small system and confidently extrapolate the results to larger systems with many non-nearest-neighbor spectator qubits. We therefore consider the low-interaction regime particularly well-suited for scaling to a large, interleaved lattice of fluxoniums and transmons. This dual-species lattice architecture offers a promising route to implementing the surface code~\cite{bravyi1998quantum, fowler2012surface, Barends2014superconducting}, with high-coherence fluxonium data qubits~\cite{nguyen2019high, somoroff2023millisecond}.

Previous work~\cite{ding2023high} on FTF system demonstrated high-fidelity CZ gates but required strong fluxonium-fluxonium capacitive coupling. Although a clever choice of coupling constants guided by perturbation theory helped achieve kilohertz-level $ZZ$ interaction in that device, this approach relies on operating in a high-coupling regime. Such a regime becomes difficult to maintain in larger systems, where uniformly low $ZZ$ interaction between all qubits is required. A natural solution is to choose lower qubit coupling strengths and eliminate non-nearest-neighbor interactions. Rosenfeld et al.~\cite{rosenfeld2024high} have proposed a resonator-coupled fluxonium device with similar motivation. Our approach addresses this limitation by using a transmon as an ancillary qubit to facilitate gates and readout, altogether avoiding the need for direct fluxonium-fluxonium coupling. Any residual fluxonium-fluxonium coupling is an unwanted effect, causing hybridization of fluxonium states and acting as a channel for crosstalk errors during gate operations. To mitigate these issues, we keep both the fluxonium-fluxonium and fluxonium-transmon couplings low. Specifically, we maintain a low fluxonium-transmon coupling to keep the $ZZ$ interaction well below the megahertz level, all while still allowing for fast gates. The CR scheme presented here offers a clearer path to scalability.

We intentionally focus on experimentally feasible approaches without the use of complicated pulse shapes. We highlight three central applications of the cross-resonance effect in our FTF system: fluxonium readout via the transmon, logical fluxonium-fluxonium gates, and $Z$-basis parity extraction of two fluxoniums. By using single-qubit phase rotations, these parity checks can be easily extended to any basis. In an experimental setting, neighboring transmons can serve as ancillas coupled to readout resonators to perform state measurements for both parity checks and general fluxonium readout. This approach simplifies the system by limiting the number of qubits directly connected to a resonator. We do not include a detailed analysis of the transmon readout itself, as transmon-resonator systems have been extensively studied in recent literature~\cite{shillito2022dynamics, cohen2023reminiscence, chen2023transmon, swiadek2024enhancing, kurilovich2025high, spring2025fast}.

The paper is structured as follows. We begin in Section~\ref{sec:II} by formulating and characterizing the energy spectra of the FT and FTF systems. In Section~\ref{sec:III}, we introduce the selective darkening cross-resonance method and provide a detailed error budget for our high-fidelity CR \textsc{cnot} gates. Section~\ref{sec:IV} describes mapping a fluxonium state to the transmon for readout, performing parity measurements, and implementing logical fluxonium-fluxonium \textsc{cnot} gates. Finally, we offer our concluding remarks in Section~\ref{sec:V}.

\section{Coupled Fluxonium-Transmon systems}
\label{sec:II}
\subsection{Static Hamiltonian}
\begin{table*}[ht]
\centering
\begin{tabular}{||c | c c c c c || c ||}%
 \hline
 Qubit & $E_J/h$ (GHz) & $E_C/h$ (GHz) & $E_L/h$ (GHz) & $f_{01}$ (GHz) & $f_{12}$ (GHz) & Coupling (MHz) \\ [0.5ex] 
 
 \hline\hline
 F1 & 3.95 & 1.4 & 0.9 & 0.88 & 3.98 & $J_1/h = 22$\\
 \hline
 T & 18 & 0.25 & -- & 5.74 & 5.46 & $J_2/h = 22$\\
 \hline
 F2 & 4.05 & 1.4 & 0.9 & 0.85 & 4.04 & $I/h = 0$\\
 \hline
\end{tabular}

\caption{System parameters}
\label{tab:tab1}
\end{table*}

\begin{figure}
\includegraphics[width=0.475\textwidth]{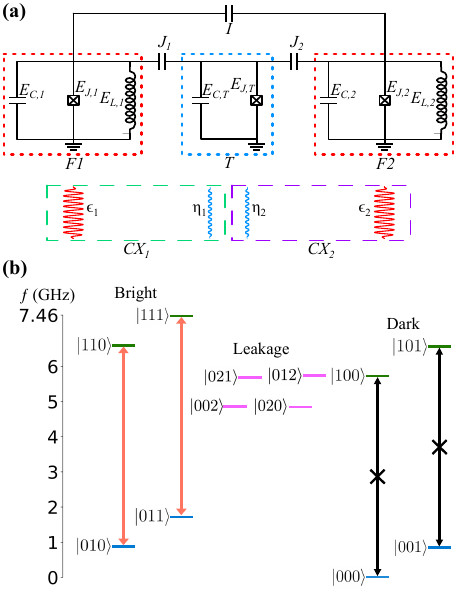}
\caption{(a): Circuit diagram of the fluxonium-transmon-fluxonium system. $\epsilon_{\alpha}$ is a scaling parameter of microwave drives with frequency $\omega_d$ coupled to F$\alpha$ and T. $\eta_{\alpha}$ defines the ratio of the same two drive amplitudes to achieve the selective darkening. (b): Eigenenergies of the three-qubit system labeled by their corresponding eigenstates with indices in the order T, F1, F2. The orange arrows show targeted transitions for which we expect a $\pi$ rotation in a $CX_1$ gate. Black arrows mark selectively darkened transitions in the same gate.}\label{fig:fig1}
\end{figure}

We consider two systems composed of capacitively coupled fluxonium and transmon qubits interacting with external drive fields. Fig.~\ref{fig:fig1} shows the circuit layout of our proposed FTF device, along with a visualization of our CR pulses, and the full dressed energy diagram. The FT system would be obtained by fully decoupling F2 from the rest of the qubits. Thus, these systems are described by Hamiltonians of the general form
\begin{align}
\label{eq:model}
    \hat{H}_{\rm FTF} = &\sum_{\alpha=F1,F2} \hat{H}_{\alpha} + \hat{H}_{T} + 
    \sum_{\alpha=F1,F2} J_{\alpha}\hat{n}_{\alpha}\hat{n}_{T}
    + I\hat{n}_{F1}\hat{n}_{F2}.
\end{align}
Individual fluxoniums F$\alpha$ are modeled by the Hamiltonian
\begin{align}
    \label{eq:H_F1}
    \hat{H}_{\alpha} & = 4E_{C,\alpha}\hat{n}_{\alpha}^2 + \frac{1}{2}E_{L,\alpha}\hat{\varphi}_{\alpha}^2 - E_{J,\alpha}\cos(\hat{\varphi}_{\alpha} - \phi_{ext,\alpha}),
\end{align}
while the transmon Hamiltonian is
\begin{align}
    \hat{H}_{T} & = 4E_{C,T}\hat{n}^2_{T} - E_{J,T}\cos\hat{\varphi}_{T}.
\end{align}
Here, $E_{C,\alpha}$, $E_{J,\alpha}$, $E_{L,\alpha}$ are the charging, Josephson, and inductive energies of individual qubits ($\alpha = F1$, $F2$, or T). For qubit $\alpha$, $\hat{n}_\alpha$ ($\hat{\varphi}_\alpha$) is the charge (flux) operator. The commutation relation $[\hat{n}_\alpha, \hat{\varphi}_\beta] = i\delta_{\alpha\beta}$ of $\hat{n}_\alpha$, and $\hat{\varphi}_\alpha$, makes them analogous to momentum and position, respectively. $J_{\alpha}$ values are the capacitive coupling constants facilitating interactions between fluxonium F$\alpha$ and the transmon, and $I$ is the fluxonium-fluxonium coupling strength. $\phi_{ext,\alpha} = 2\pi \Phi_\alpha/\Phi_0$ is the external flux through the qubit's superconducting loop normalized by the flux quantum $\Phi_0$, to be dimensionless. Fluxonium qubits in our numerical analysis are always in the flux ``sweet spot'' of maximal coherence time at $\phi_{ext,\alpha} = \pi$. We label dressed states of the FT system as $\ket{ij}$ where $i$ ($j$) corresponds to the index of the $\rm{F1 (T)}$ eigenstate. For the FTF system, we label dressed states as $\ket{ijk}$ where $i$ indexes the transmon eigenstates, $j$ ($k$) indexes the eigenstates of $\rm{F1 (F2)}$.

\subsection{Energy Spectrum and $ZZ$ Interactions}

Unless otherwise specified, the numerical values of system parameters are provided in Table~\ref{tab:tab1}. These parameters characterize a system with the energy spectrum shown in Fig.~\ref{fig:fig1}. Our qubit parameters are chosen to have the necessary large cross matrix elements and elements and suppression of control qubit flips by detuning. They also reflect physically feasible and noise insensitive regimes, but the parameters themselves are not unique in their high-performance.
\begin{figure}
\includegraphics[width=0.475\textwidth]{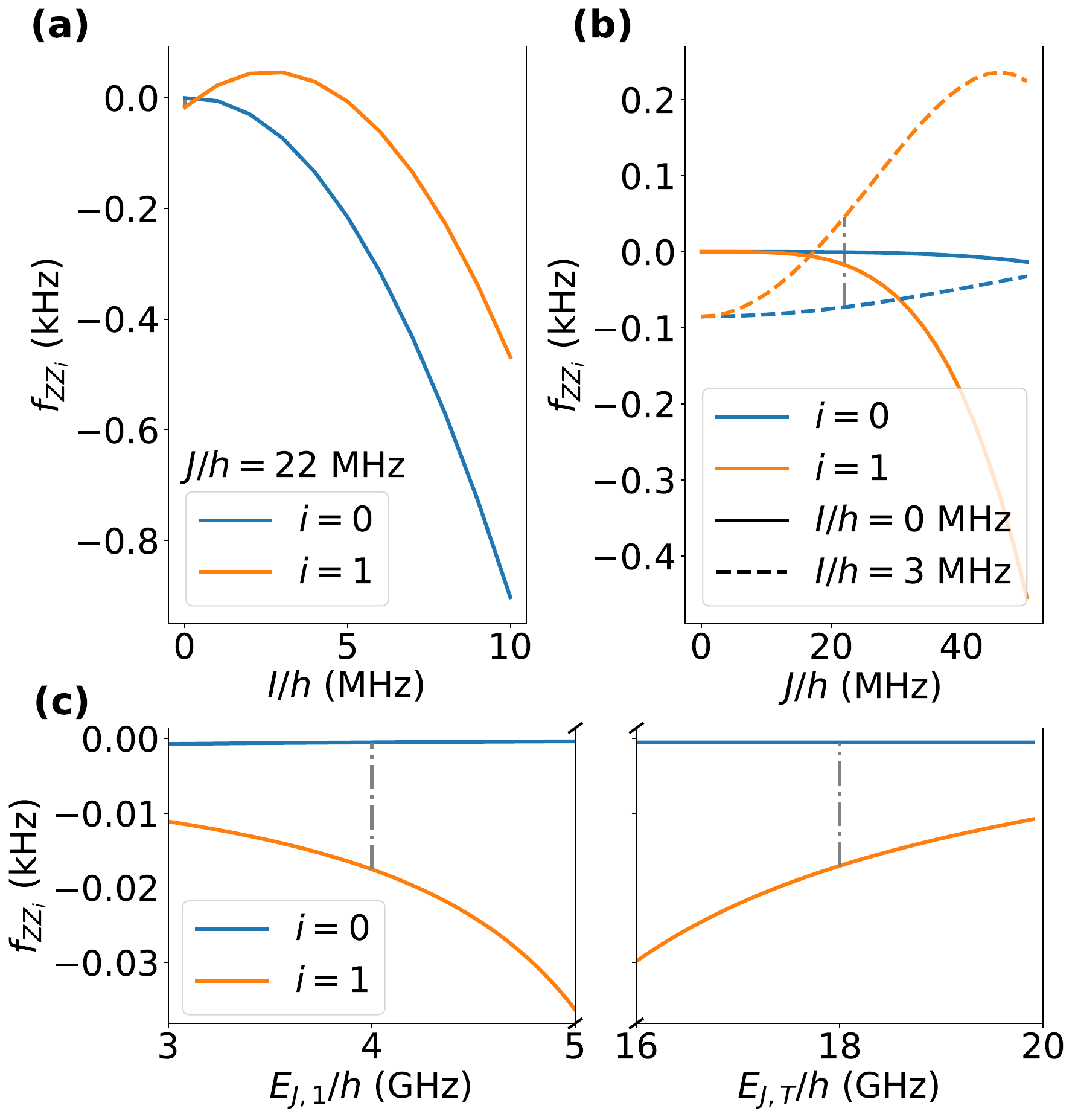}
\caption{Fluxonium-fluxonium $ZZ$ interaction strength as a function of (a-b) qubit coupling parameters, and (c) Josephson energies. Due to the equality of F1 and F2 parameters other than $E_J$, $E_{J,1}$ effectively also represents $E_{J,2}$. In all plots, $i$ denotes the transmon's state. $f_{ZZ_i}$ is plotted for values of $i = 0,1$ to demonstrate the range of data qubit $ZZ$ dependence on ancilla qubit computational states. Vertical grey lines highlight this variance at each parameter's point of interest for the rest of the paper. The seemingly straight blue lines in subplot (c) have values on the order of Hertz.}\label{fig:fig2}
\end{figure}
From the energy spectrum we can extract always-on $ZZ$ entanglement between qubit pairs. In the full 3-qubit system, the $ZZ$ interaction strength between fluxonium qubits also depends on the state of the transmon. Given our proposal of using fluxoniums as data qubits, this value can limit performance and is therefore of interest. For T in state $i$, fluxonium-fluxonium $ZZ$ rate is given by
\begin{equation}
    f_{ZZ_i} = E_{i11} + E_{i00} - E_{i10} - E_{i01}.
\end{equation}

Our system with parameters in Table~\ref{tab:tab1} has fluxonium-fluxonium $ZZ$ values of $f_{ZZ_0} = -0.5$ Hz and $f_{ZZ_1} = -17$ Hz. The dependence of these unwanted effects on varied system parameters is shown in Fig.~\ref{fig:fig2}. These parasitic coupling strengths stay well below the kilohertz-level as a consequence of small or zero direct fluxonium-fluxonium coupling.

Along with keeping $ZZ$ interaction strengths between fluxoniums low, their values should be insensitive to the transmon state. The exact trends for each $Z$-basis state of T are shown in Fig.~\ref{fig:fig2} with respect to different parameters. By targeting negligible $I$ values, $f_{ZZ_i}$ stays well below the kilohertz level, and therefore $ZZ$ cancellation pulses~\cite{xiong2022arbitrary} are not necessary for an effective gate in our FTF system.

Along with keeping $ZZ$ interaction strengths between fluxoniums low, their values should be insensitive to the transmon state. The exact trends for each $Z$-basis state of T are shown in Fig.~\ref{fig:fig2} with respect to different parameters. By targeting negligible $I$ values, $f_{ZZ_i}$ stays well below the kilohertz level, and therefore $ZZ$ cancellation pulses~\cite{xiong2022arbitrary} are not necessary for an effective gate in our FTF system.

\begin{figure}
\includegraphics[width=0.475\textwidth]{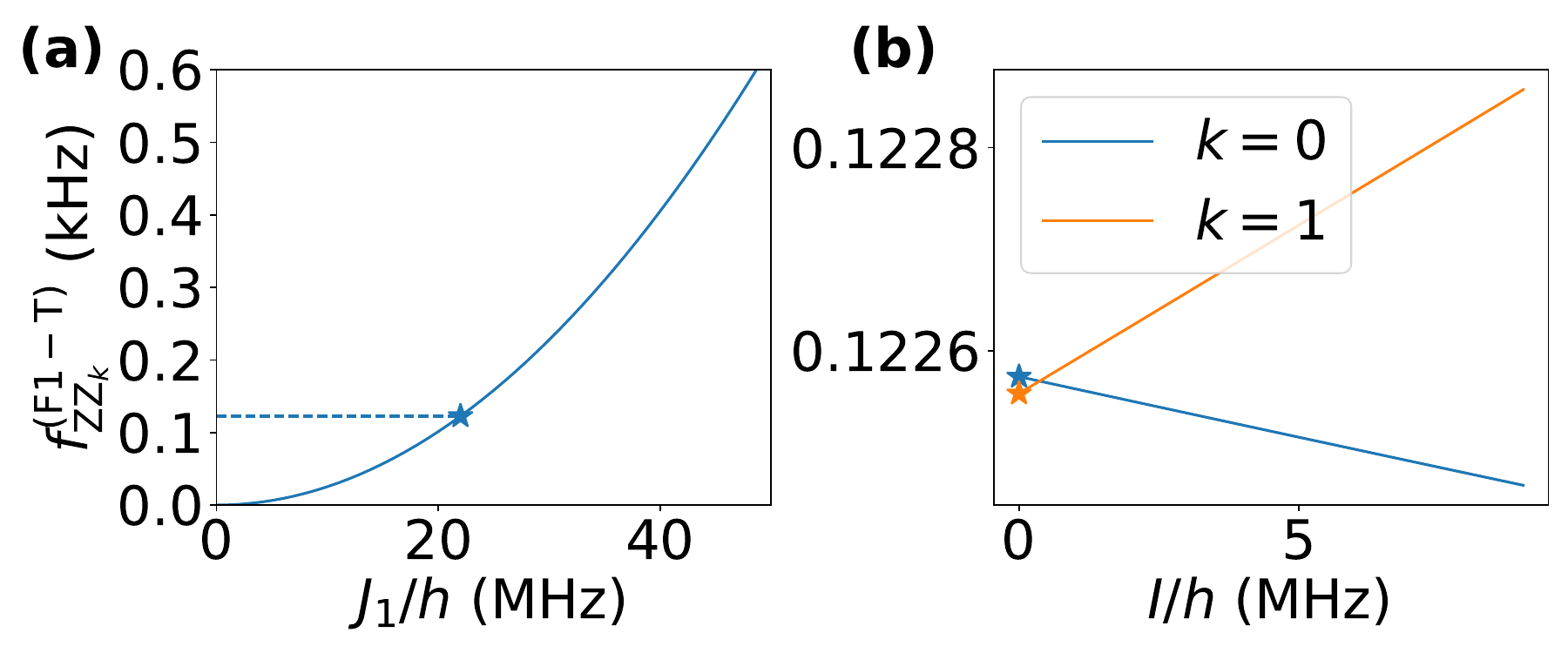}
\caption{$ZZ$ coupling between F1 and T as a function of (a) coupling constant $J/h$ when $I/h = 0$ and (b) $I/h$ when $J/h = 22$ MHz for F2 in state $k$. In (a), all computational states of F2 give indistinguishable $ZZ$ values within the shown visual accuracy. (b) demonstrates low sensitivity of the fluxonium-transmon $ZZ$ coupling on $I$.}\label{fig:fig3}
\end{figure}

Since we use our transmon as an ancillary qubit, fluxonium-transmon $ZZ$, is only important through its impact on gates. The dependence of this quantity on coupling strengths is shown in Fig.~\ref{fig:fig3}, and it is defined as
\begin{equation}
    f_{ZZ}^{\rm{(F\alpha-T)}} = E_{11i} + E_{00i} - E_{01i} - E_{10i},
\end{equation}
when indexing as $\ket{T,F\alpha,F\beta}$.
During \textsc{cnot} gates in the three-qubit system, these interactions cause the targeted transmon transition to shift based on spectator states. As elaborated upon in Sec.~\ref{sec:3qcxerrorbudget}, this mechanism sets the minimum increase in gate error when adding a spectator. Despite this, our choice of coupling constants ensures that a longer chain of interleaved fluxoniums and transmons would not experience significantly stronger $ZZ$ effects. Additionally, the induced shift on energy levels for a transmon in a lattice of nearest-neighbor connected fluxoniums and transmons can be easily extrapolated for gate performance in that architecture.
\section{Selective darkening \textsc{cnot} gate}
\label{sec:III}
Before discussing hardware-specific implementations, let us consider the dynamics of a CR gate on two ideal qubits. By driving the control close to the target qubit's $\ket{0}\leftrightarrow\ket{1}$ transition frequency, we can achieve an entangling gate via the cross-resonance effect. To demonstrate this, we use a simplified model of transversally coupled qubits labeled $A$ (control) and $B$ (target). As such, we start with the driven interacting Hamiltonian
\begin{equation}
\hat{H} = \frac{\omega_A}{2} \sigma_z^A + \frac{\omega_B}{2} \sigma_z^B 
+ J \sigma_x^A \sigma_x^B 
+ \Omega_d \cos(\omega_B t) \sigma_x^A.
\end{equation}

We then transform to a frame rotating with the qubits, and apply the rotating-wave approximation to get
\begin{align}
\hat{H}_{\text{rot}}(t) &= 
J \left( \sigma_+^A \sigma_-^B e^{-i\Delta t} + \sigma_-^A \sigma_+^B e^{i\Delta t} \right) \nonumber\\
&\quad + \frac{\Omega_d}{2} \left( \sigma_+^A e^{-i\Delta t} + \sigma_-^A e^{i\Delta t} \right),
\label{eq:Hrot}
\end{align}
where $\Delta = \omega_A - \omega_B$. Treating $J$ and $\Omega_d$ as small parameters compared to $|\Delta|$, we apply a Schrieffer--Wolff transformation and obtain the second-order static Hamiltonian
\begin{equation}
\hat{H}_{\rm eff} \approx \frac{J\Omega_d}{\Delta} \sigma_z^A \sigma_x^B
+ \frac{J^2}{\Delta} \sigma_z^A \sigma_z^B
+ \frac{\Omega_d^2}{4\Delta} I \otimes \sigma_x^B.
\label{eq:Heff}
\end{equation}
\begin{itemize}
\item The $\sigma_z^A\sigma_x^B$ (\textbf{ZX}) term is the desired entangling interaction that enables an effective \textsc{cnot} operation.
\item The $\sigma_z^A\sigma_z^B$ (\textbf{ZZ}) term represents an unwanted longitudinal coupling of the qubits. This contribution can be eliminated with single-qubit phase rotations before and after the gate.
\item The $I\otimes\sigma_x^B$ (\textbf{IX}) term represents an effective direct drive on $B$ due to off-resonant excitation of $A$ and coupling through $J$. This term can be canceled or enhanced with an additional target qubit drive.
\end{itemize}

For selective darkening~\cite{de2010selective, deGroot2012SD}, the \textbf{IX} and \textbf{ZX} rates ideally have the same amplitude to achieve the correct interference. Methods for doing so are hardware dependent and will be further discussed for the charge-driven FT and FTF systems.
\subsection{FT system}
\label{sec:2qubit}
As a building block, we now consider the performance of the CR \textsc{cnot} gate in a system of one fluxonium coupled to a transmon \textit{i.e.} $J_{2}=0$ and $I=0$.
Following Eq.~\eqref{eq:model}, with the addition of a microwave drive field, we write the Hamiltonian
\begin{align}
\label{eq:2qubitH}
    \hat{H}_{\rm FT}(t) = \hat{H}_{\rm F,1} + \hat{H}_{\rm T} + J\hat{n}_{\rm F1}\hat{n}_{\rm T} + \hat{H}_{\rm dr}.
\end{align}
Charge-coupled microwave gate pulses are represented by
\begin{align}
\label{eq:2qdrive}
\hat{H}_{\rm dr} = \epsilon f(t)\cos(\omega_{d} t) (\hat{n}_{\rm F1} + \eta \hat{n}_{\rm T}).
\end{align}
In principle, flux-driving fluxoniums is also possible, but we consider the fully charge-driven case for uniformity. As with all future gates, we assume that microwave pulses are applied simultaneously to the control and target qubits without any timing delays. Both drive amplitudes scale proportionally to $\epsilon$, and their ratio is set by the coefficient $\eta$. F$\alpha$ is the control qubit, T is the target qubit, and $f(t)$ is an envelope function described by
\begin{align}
\label{eq:envelope}
f(t) =
\begin{cases}
\sin^2 \left( \frac{\pi t}{2 t_{\rm r}} \right) & \text{if } t < t_{\rm r}, \\
1 & \text{if } t_{\rm r} \leq t < t_{\rm g} - t_{\rm r}, \\
\sin^2 \left( \frac{\pi (t_{\rm g} - t)}{2 t_{\rm r}} \right) & \text{if } t \geq t_{\rm g} - t_{\rm r}.
\end{cases}
\end{align}
We construct a \textsc{cnot} gate between qubits F1 (control) and T (target) using a CR microwave pulse of both qubits at the frequency $\omega_{d}$. Although the transitions $\ket{10}\leftrightarrow\ket{11}$ and $\ket{00}\leftrightarrow\ket{01}$ are close in frequency, we aim to enhance the former while suppressing the latter. Each qubit's coupled drive amplitude is chosen to achieve selective darkening of the unwanted transition, as described by the conditions
\begin{subequations}
\label{eq:sdconditions}
\begin{align}
\bra{01}\hat{H}_{\rm dr}\ket{00} = 0, \\
\bra{10}\hat{H}_{\rm dr}\ket{11} \ne 0.
\end{align}
\end{subequations}
The perturbative result 
\begin{equation}
    \eta = -\bra{00}\hat{n}_{F1}\ket{01}/\bra{00}\hat{n}_{T}\ket{01}
\end{equation}
provides an approximate value of the optimal drive amplitude ratio. Optimization around that value is computationally cheap due to the smooth parabolic dependence of total error on $\eta$. For system parameters in Table~\ref{tab:tab1}, $\eta \approx 1.01 \cdot 10^{-3}$ satisfies Eq.~\eqref{eq:sdconditions}, while $\eta \approx 1.36 \cdot 10^{-3}$ maximizes fidelity for a 50 ns gate.

The ideal \textsc{cnot} gate can be expressed in the computational basis as
\begin{align}
\label{eq:2qcx}
\hat{U}_{\rm id} = \begin{pmatrix}
1 & 0 & 0 & 0\\
0 & 1 & 0 & 0\\
0 & 0 & 0 & 1\\
0 & 0 & 1 & 0
\end{pmatrix}.
\end{align}
We simulate the driven evolution operator as $\hat{U} = \hat{\mathcal{T}}\exp{(-i\int_0^{t_g}{\hat{H}_{\rm FT}(t)dt})}$ with the time-ordering operator $\hat{\mathcal{T}}$. Truncation of each qubit to its lowest 6 energy levels is conducted before numerical integration. This value was determined to be sufficient by examining matrix elements and transition detunings compared to drive frequency. Coherent fidelity $\mathcal{F}_{\rm coh}$ is then evaluated with respect to $U_{\rm id}$ using the process fidelity expression 
\begin{equation}
\label{eq:fidelity}
\mathcal{F}_{\rm coh}(\hat{U}, \hat{U}_{id}) = \frac{\mathrm{Tr}(\hat{U}^{\dag}\hat{U}) + \left|\mathrm{Tr}(\hat{U}_{\rm id}^{\dag}\hat{U})\right|^2}{d(d+1)}.
\end{equation}
Here, $d = 4$ for the two-qubit system. Total gate error is then calculated as $\mathcal{E} = 1 - \mathcal{F}_{\rm coh}$. Before computing fidelity, we correct for qubit phase accumulation differences following the procedure established by Nesterov et al.~\cite{nesterov2022cnot}.
\begin{figure}
\includegraphics[width=0.475\textwidth]{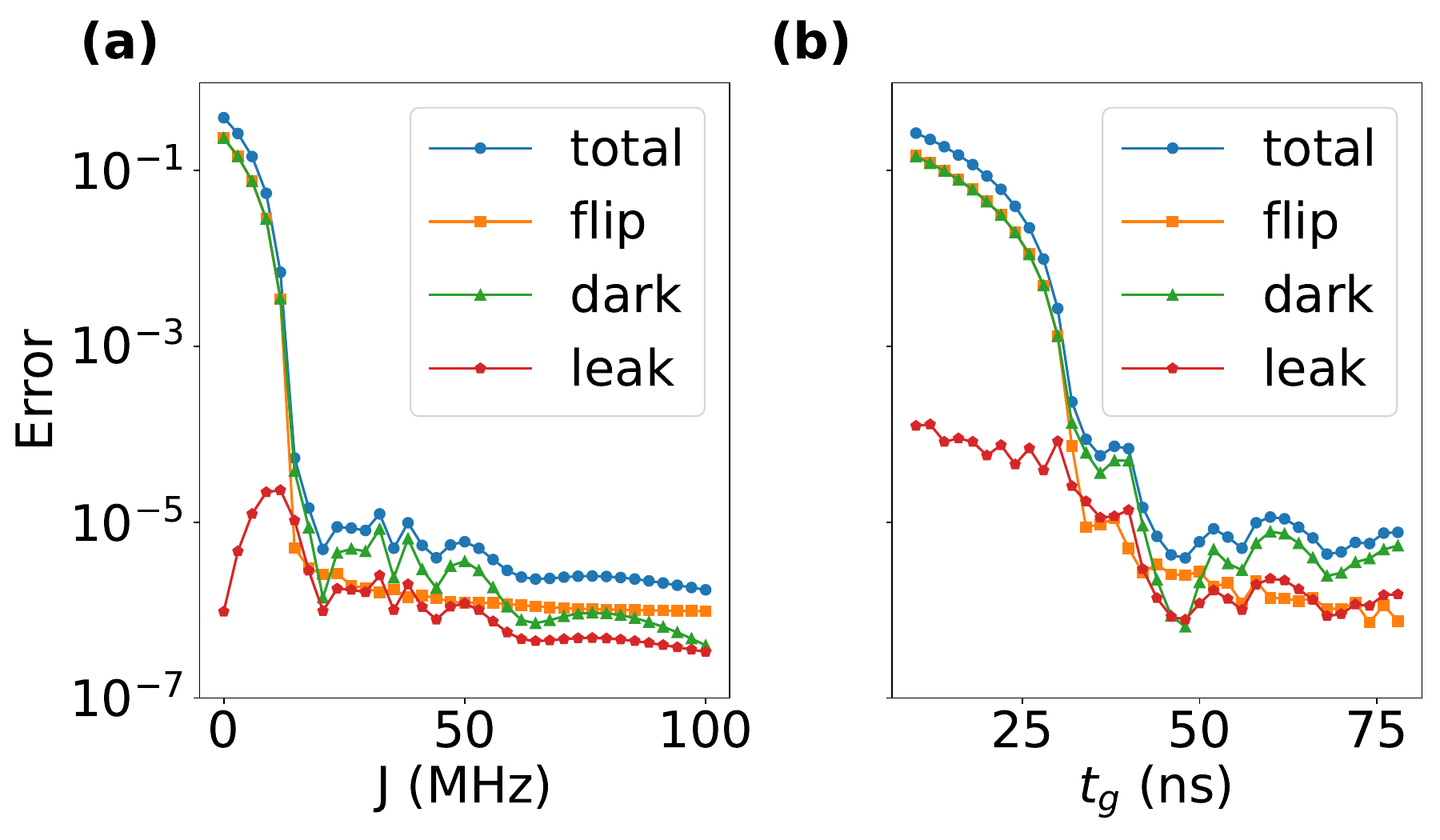}
\caption{ Optimized gate error as a function of (a) coupling constant $J/h$ when $t_g$ is fixed to 50 ns and (b) gate time $t_g$ when $J/h$ is fixed to $22$ MHz for a \textsc{cnot} gate in the 2-qubit system. Visible oscillatory behavior is the result of varied timing of counter-rotating terms for optimal drive parameters.}\label{fig:fig4}
\end{figure}
Given correctly chosen drive parameters, transitions involving F1 in the ground state remain dark while $\ket{10}$ and $\ket{11}$ undergo Rabi flips of the transmon. For all gate pulses, we set the carrier rise time to $t_{\rm r} = 5$ ns for effective adiabatic switching of the drive field strength. By numerical minimization of $\mathcal{E}$ with control parameters $\epsilon$, $\eta$, and $\omega_{d}$, we obtain gates with error on the order of $10^{-5}$.

In previous work for fluxonium-fluxonium devices, an error budget for CR \textsc{cnot} gates was formulated by grouping errors of similar nature and magnitude into 5 categories~\cite{nesterov2022cnot}. Although the same approach applies to this work, the number of error terms scales poorly with qubit count. For example, we find that our FTF system would require 13 error transition groups for full characterization in the same manner. To avoid convolution, we establish a new error budget with the four categories $\mathcal{E}_{\rm dark}, \mathcal{E}_{\rm flip}, \mathcal{E}_{\rm leak}, \mathcal{E}_{\rm phase}$ regardless of system size. In the absence of spectators, as is the case for the FT system, $\mathcal{E}_{\rm phase} = 0$, so we only have three relevant categories. $\mathcal{E}_{\rm dark}$ contains error contributions of transitions from a selectively darkened \textit{i.e.} ``dark'' state. $\mathcal{E}_{\rm flip}$ represents incorrect transitions from ``bright'' states on which the evolution operator should execute a transmon $\pi$ rotation. $\mathcal{E}_{\rm leak}$ captures error contributions associated with non-unitary evolution. To express these terms, we define shorthand notation $\Pi_{ab\to a'b'} = \bra{a'b'}\hat{U}_{\rm sim}\ket{ab}$, and $\phi_{ab}$ such that $\Pi_{ab\to a'b'} = \exp(i \phi_{ab})|\Pi_{ab\to a'b'}|$ for matrix elements with an ideal magnitude of 1. Ignoring the global phase, all phase factors in numerical simulations are defined to follow $\phi_{ab} = \phi_{ab} - \phi_{00}$.
Then, decomposing $\mathcal{E}$ following Appendix~\ref{appendix:errorbudget} leads to the terms
\begin{subequations}
    \begin{equation}
        \mathcal{E}_{\rm dark} = \frac{2}{5} \sum_{i=0}^1 (1 - \Pi_{0i \to 0i}),
    \end{equation}
    \begin{equation}
        \mathcal{E}_{\rm flip} = \frac{2}{5} \sum_{i=0}^1 (1 - \Pi_{1i \to 1\bar{i}}),
    \end{equation}
    \begin{equation}
        \mathcal{E}_{\rm leak} = 1/5 - {\mathrm{Tr} }\{\hat{U}_{\rm sim}^{\dagger} \hat{U}_{\rm sim}\}/20.
    \end{equation}
    \label{eq:2qerrorterms}
\end{subequations}
We treat the fluxonium-transmon system as a probe for optimal performance, assuming full isolation of the two-qubit system. In later sections with an additional fluxonium, we can compare the results with the two-qubit case as a baseline. This comparison allows us to effectively isolate the effect of a spectator qubit coupled to the system. It is important to note that all results we present assume full qubit coherence. For each active qubit with known relaxation and pure dephasing times $T_1$ and $T_{\phi}$ respectively, its additional error contribution can be estimated as $\mathcal{E}_{\rm de} \approx 1 - \exp[-t_g (1/T_1 + 1/T_{\phi})]$. This estimate can be used in conjunction with the coherent relationship between $\mathcal{F}_{\rm coh}$ and $t_g$ to determine an experimental sweet spot for $t_g$ that maximizes gate fidelity. The optimized fidelity dependence on $J$ also shows smooth behavior with stable performance at well below $10^{-5}$ total error.
\subsection{The \textsc{cnot} gate for a system of two fluxoniums coupled to a transmon}
\label{sec:FTFCX}

\subsubsection{CX gate realization}

A natural extension of the system we consider in Sec.~\ref{sec:2qubit} is the addition of another fluxonium (F2) strongly coupled to T and either weakly coupled to or fully decoupled from F1. It follows that the Hamiltonian in Eq.~\eqref{eq:2qubitH} acquires terms for the third qubit and its interactions such that
\begin{equation}
    \hat{H}_{\rm FTF}(t) = \hat{H}_{\rm FTF}+\hat{H}_{\rm dr},
\end{equation}
where $J_2$ and $I$ are no longer necessarily zero. For uniformity, we assume that the coupling between the transmon and either fluxonium is given by the same interaction strength $J$. Direct coupling strength between fluxoniums $I$ is kept at $0$ for all computations unless otherwise specified. The microwave drive pulse for a \textsc{cnot} gate between control qubit F$\alpha$ and target qubit T is described by the term
\begin{align}
\hat{H}_{\rm dr,\alpha} = \epsilon_\alpha f(t)\cos(\omega_{d} t) (\hat{n}_{\rm \rm{F} \alpha} + \eta_{\alpha} \hat{n}_{\rm T}).
\end{align}
As with gates in the two-qubit system, pulse carrier envelopes are defined by Eq.~\eqref{eq:envelope}.

For the three-qubit system, we now use all parameters in Table~\ref{tab:tab1}. Based on the energy configuration as seen in Fig.~\ref{fig:fig1}, one possible concern with these parameters is leakage resonances. In particular, $f_{02}^{F\alpha} + f_{01}^{F\beta}$ for either order of fluxoniums $F_\alpha$ and $F_\beta$ has a value close to $f_{01}^{T}$. Despite this, our chosen coupling constant values and drive amplitudes suppress these transitions. In particular, these two seemingly resonant leakage transitions, $\ket{021}$ and $\ket{012}$ require specific excitations in both the control and spectator fluxoniums. The first transition is suppressed by high qubit anharmonicity, and the lack of direct drive strongly suppresses the second transition. This makes the only leakage pathway a shared capacitance with the transmon, which is already weakly driven. As a result, $\ket{000}\xrightarrow{}\ket{021}$ and $\ket{000}\xrightarrow{}\ket{012}$ errors during \textsc{cnot} gates are negligible compared to the dominant error transitions.

To perform a \textsc{cnot} operation with control F1 and target T, or control F2 and target T, we use the same CR approach as with the FT system, but now in the presence of capacitive interaction with the non-driven qubit. For active fluxonium F$\alpha$ and spectator fluxonium F$\beta$, the gate is denoted $CX_{\alpha}$ and has an evolution operator $U_{\rm CX_{\alpha}} = \hat{\mathcal{T}}\exp{(-i\int_0^{t_g}{\hat{H}_{\rm FTF}(t)dt})}$. The selective darkening condition is now
\begin{align}
    \label{eq:3qSDcondition}
    \eta_{\alpha |\beta} = -\frac{\bra{0_T0_{\alpha}\beta}\hat{n}_{\alpha}\ket{1_T0_{\alpha}\beta}}{\bra{0_T0_{\alpha}\beta}\hat{n}_{T}\ket{1_T0_{\alpha}\beta}},
\end{align}
where indexing is performed in the order of transmon, active fluxonium F$\alpha$, spectator fluxonium F$\beta$. The dependence of $\eta_{\alpha |\beta}$ on $\beta$ is the determining factor for the feasibility of spectator-state-independent selective darkening. Errors of this nature are later categorized as ``dark'' errors, and they are not dominant in the low-error regime. Further, we see that the dominant errors are in fact associated with the choice of drive frequency rather than drive amplitudes. The ideal gates for $CX_1$ and $CX_2$ are
\begin{align}
\hat{U}_{\rm CX_1} = \hat{U}_{\rm F1,T}\otimes\hat{I}_{\rm F2},\\
\hat{U}_{\rm CX_2} = \hat{I}_{\rm F1}\otimes\hat{U}_{\rm F2,T}
\end{align}
respectively, where $\hat{U}_{\rm \rm{F} \alpha,T}$ is a \textsc{cnot} gate between F$\alpha$ (control) and T (target). Due to our indexing order of T, F1, F2, we use the symbol ``$\otimes$'' to generally denote a tensor product of Hilbert spaces but not a Kronecker product of matrices with a specific order of indexing. As with the FT system, we calculate the gate fidelity $\mathcal{F}_{\rm coh}$ using Eq.~\eqref{eq:fidelity}, but now with $d = 8$ for our $8 \times8$ matrices.
\begin{figure*}[ht]
\centering
\includegraphics[width=1\textwidth]{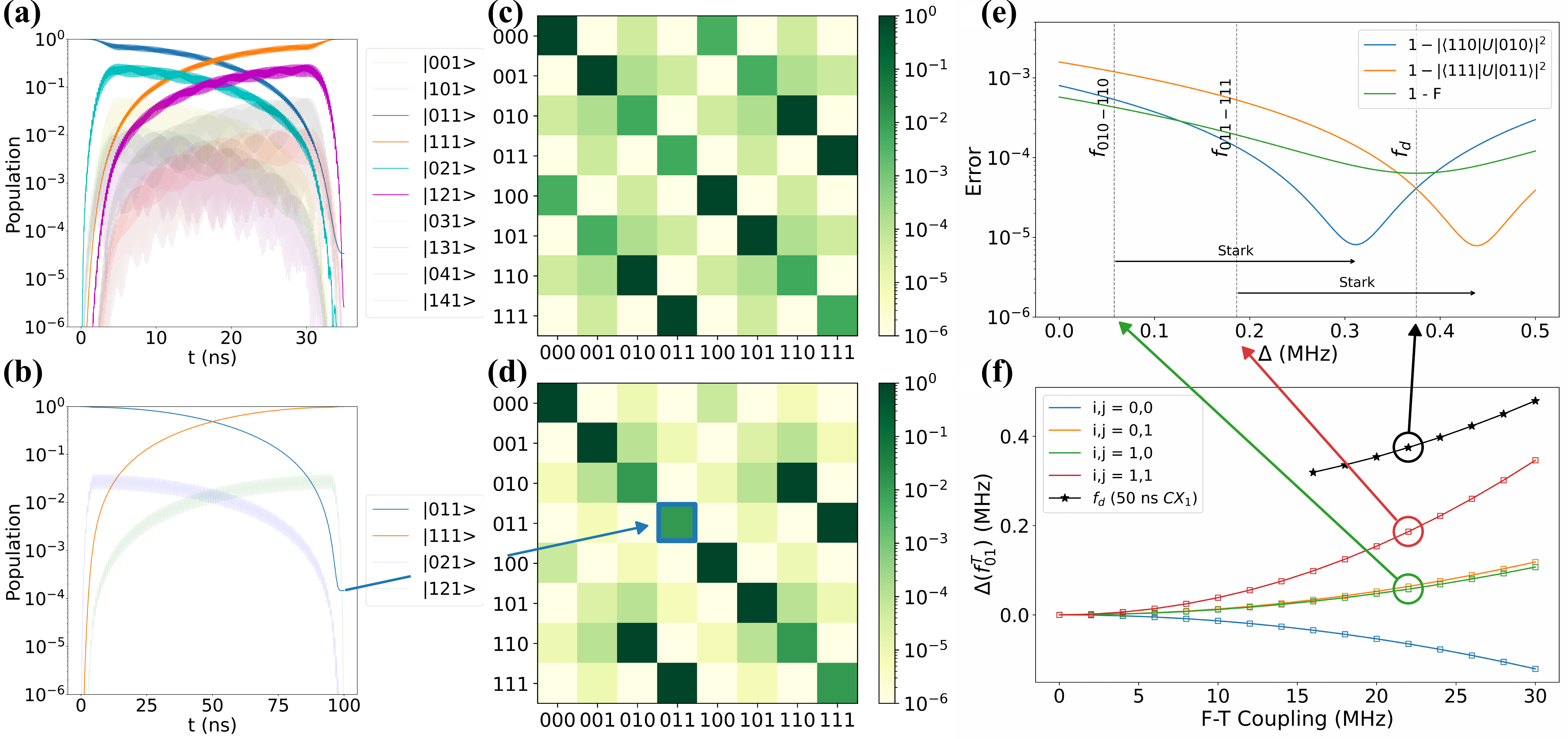}
\caption{(a-b): Eigenstate population dynamics of initial state $\ket{011}$ during $CX_1$ gates optimized for pulse times of (a) 35 ns, and (b) 100 ns. The gates achieve respective fidelities of (a) $99.989\%$, (b) $99.975\%$. Only eigenstates that achieve a population of at least $10^{-2}$ at any point during the pulse are plotted and those that reach $10^{-1}$ are shown in full intensity for visual distinction. (c-d): Evolution operator diagrams of $|\hat{U}|$ gates with pulse times of (c) 35 ns and (d) 100 ns. (e): 50 ns $CX_1$ gate error contributions as a function of drive detuning $\Delta = \omega_d - f_{01}^T$. The optimized value balances error caused by spectator state-dependent detuning after accounting for drive-induced AC-Stark shifts. (f): $\Delta(f_{01}^T) = \tilde{f_{01}^T} - f_{01}^T$ as a function of $J/h$ where $\tilde{f_{01}^T}$ is dressed by $\rm{F1 (F2)}$ in state $i$ ($j$) and $f_{01}^T$ is the bare value. Square markers show results from second order perturbation theory while ignoring terms with charge matrix elements $<0.01$. Black stars show $f_d - f_{01}^T$ where $f_d$ values have been chosen by the optimizer for a gate with $t_g = 50$ ns.}
    \label{fig:fig5}
\end{figure*}
 As seen in Table~\ref{tab:tab1}, we choose close to identical parameters for the two fluxoniums. The variance in $E_J$ values avoids degeneracy and reflects a small tolerance in fabrication. This approach highlights our system's insensitivity to resonances between transitions in different fluxoniums. As a result of symmetric parameter choices, we mostly present data for $CX_1$ gates. Even so, all concepts also apply to $CX_2$ gates with similar respective numerical values. With current fabrication techniques, qubit $E_J$ values have non-negligible variance. To somewhat emulate this effect and avoid degeneracy in numerical diagonalization, we choose $E_{J,1}$ and $E_{J,2}$ values with differences of $-0.05$ GHz and $0.05$ GHz, respectively, from the ``targeted'' value of $4$ GHz. Variance in this parameter is case-specific and often higher. As such, we further investigate its effect in Appendix~\ref{sec:EJStability}. Due to the system's insensitivity to crosstalk effects of next-nearest-neighbor qubit frequency collisions, fluxonium $E_J$ variance can be arbitrarily small. For this reason, we envision straightforward scaling to an interleaved fluxonium-transmon lattice by targeting the same parameters for all fluxoniums. We use $E_{ijk}$ to denote the eigenenergy associated with eigenstate $\ket{i_Tj_{\rm{F} \alpha}k_{\rm{F} \beta}}$. Due to the dependence of optimal $\omega_{d}$ on the spectator qubit state, we now optimize around $\bar{\omega}_{\rm d} = (E_{110} + E_{111} - E_{010} - E_{011})/2$. By minimizing $\mathcal{E}$ for $t_{\rm g} = 50$ ns with control parameters $\epsilon_\alpha$ and $\eta_\alpha$, we obtain fidelity values above $99.994\%$ for both the $CX_1$ and the $CX_2$ gate. The primary reason for a drop in fidelity compared to the two-qubit case comes from the spectator qubit's shift on the dressed transmon frequency, as shown in Fig.~\ref{fig:fig5}. The resulting spectator-state dependent $ZZ$ between active qubits cannot be fully canceled with single-qubit gates. As a result, remaining phase accumulation is quantified by the ``phase'' terms of our error budget discussed in Sec.~\ref{sec:3qcxerrorbudget}. It is worth noting that in a scaled-up system using our architecture, these ``phase'' errors degrade the performance of concurrent gates involving neighboring qubits. To mitigate these, one can either employ AC-Stark shift drives to cancel $ZZ$ or plan gates in such a way that avoids problematic spatio-temporal collisions.

One subtlety in our numerical results is that we perform optimization in two steps. The first step varies $\epsilon$, $\eta$, and $\omega_d$ to minimize the error of a modified evolution operator that has no phase components compared to the ideal \textsc{cnot} gate operator. The second step applies single-qubit phase rotations before and after the actual evolution operator to minimize the true error. As mentioned above, spectator-state dependent $ZZ$ of the active qubits establishes a limit on the performance difference between the two steps. Due to the insensitivity of this $ZZ$ strength to $\epsilon$, $\eta$, and $\omega_d$, ignoring it during the first optimization step is a valid approach. Thus, we conclude that our two-step optimization is effectively equivalent to optimizing all parameters together, but computationally more efficient. For a two-qubit system, the first step is the only necessary optimization, because local phase rotations can be found analytically to fully eliminate the effect of $ZZ$.

Keeping $ZZ$ low allows the optimizer to find a set of drive parameters closer to satisfying the selective darkening condition in the presence of an arbitrary computational spectator state. This can be achieved by lowering the coupling constants $J_{1}, J_2$. We do so uniformly with $J$ to keep the performance of $CX_1$ and $CX_2$ as close as possible. However, having low coupling constants requires an increase in drive amplitude to perform the same rotations on the Bloch sphere. This requirement can introduce a new dominant error in the form of excitation to leakage states. Longer gate times would resolve this issue by enabling lower drive amplitudes and less leakage. We purposely keep gate times fast for experimental feasibility, where qubit decoherence must be considered.

\subsubsection{CX error budget}
\label{sec:3qcxerrorbudget}

Given that the number of computational matrix elements is now 64, individually tracking and grouping each of the 56 error transitions is especially uninsightful. Instead, we once again construct an error budget around the matrix elements ideally equal to unity. Following this approach, we define total error on an optimized gate as
\begin{align}
    \mathcal{E} = \mathcal{E}_{\rm dark} + \mathcal{E}_{\rm flip} + \mathcal{E}_{\rm leak} +
    \mathcal{E}_{\rm phase}.
\end{align}
In this decomposition, $\mathcal{E}_{\rm dark}$ represents gate errors when starting from an initial state with F$\alpha$ in state $\ket{0}$. All transitions should be dark from such initial states, so excitations in any qubit contribute to this term. $\mathcal{E}_{\rm flip}$, on the other hand, corresponds to errors accumulated from initial states with active fluxonium in state $\ket{1}$. These states require a transmon flip without disturbing the state of either fluxonium. Error contributions are evaluated using matrix elements of the evolution operator,
\begin{equation}
\Pi_{ijk \to i'j'k'} = \bra{i'_Tj'_{\rm{F} \alpha}k'_{\rm{F} \beta}}\hat{U}_{\rm CX_{\alpha}}\ket{i_Tj_{\rm{F} \alpha}k_{\rm{F} \beta}}.
\end{equation}
 Phase $\phi_{abc}$ for the major matrix element of initial state $\ket{abc}$ can be defined relative to $\phi_{000}$ such that $\Pi_{abc\to a'b'c'} = \exp(i \phi_{abc})|\Pi_{abc\to a'b'c'}|$. Following the derivation in Appendix~\ref{appendix:errorbudget}, we get the error budget terms
\begin{subequations}
    \begin{equation}
        \mathcal{E}_{\rm dark} = \frac{2}{9} \sum_{i,j=0}^1 \left(1 - \left|\Pi_{i0j \to i0j}\right|\right),
    \end{equation}
    \begin{equation}
        \mathcal{E}_{\rm flip} = \frac{2}{9} \sum_{i,j=0}^1 \left(1 - \left|\Pi_{i1j \to \bar{i}1j}\right|\right),
    \end{equation}
    \label{eq:darkflipterms}
\end{subequations}
where $\bar{0} = 1$ and $\bar{1} = 0$. Additionally, we now have the term $\mathcal{E}_{\rm phase} = \mathcal{E}_{\rm d,phase} + \mathcal{E}_{\rm b,phase} + \mathcal{E}_{\rm imag}$ associated with spectator-dependent $ZZ$ between the active qubits. The phase error components are defined as
\begin{subequations}
    \begin{equation}
        \begin{split}
        \mathcal{E}_{\rm d,phase} = \frac{2}{9} \sum_{i,j=0}^1{\left|\Pi_{i0j \to i0j}\right|(1 - \cos{\phi_{i0j}})},
        \end{split}
    \end{equation}
    \begin{equation}
        \begin{split}
        \mathcal{E}_{\rm b,phase} = \frac{2}{9} \sum_{i,j=0}^1{\left|\Pi_{i1j \to \bar{i}1j}\right|(1 - \cos{\phi_{i1j}})},
        \end{split}
    \end{equation}
    \begin{equation}
        \begin{split}
        \mathcal{E}_{\rm imag} = - &\frac{1}{72}\bigg[\sum_{i,j=0}^1\big[ \left|\Pi_{i0j \to i0j}\right| \sin{\phi_{i0j}}
        \\
        +\; &\left|\Pi_{i1j \to \bar{i}1j}\right| \sin{\phi_{i1j}}\big]\bigg]^2.
        \end{split}
    \end{equation}
\end{subequations}
As such, the entire term is always positive. In the FT system, single-qubit phase correction gates fix angles to $\phi = 0$ for each matrix element. This correction necessarily enforces our statement that $\mathcal{E}_{\rm phase} = 0$ due to the lack of spectators. In a similar fashion to the two-qubit system analysis, we separate leakage errors as
\begin{equation}
\mathcal{E}_{\rm leak} = 1/9-{\mathrm{Tr} }\{\hat{U}_{\rm CX_{\alpha}}^{\dagger} \hat{U}_{\rm CX_{\alpha}}\}/72.
\end{equation} 
\begin{figure}
\includegraphics[width=0.475\textwidth]{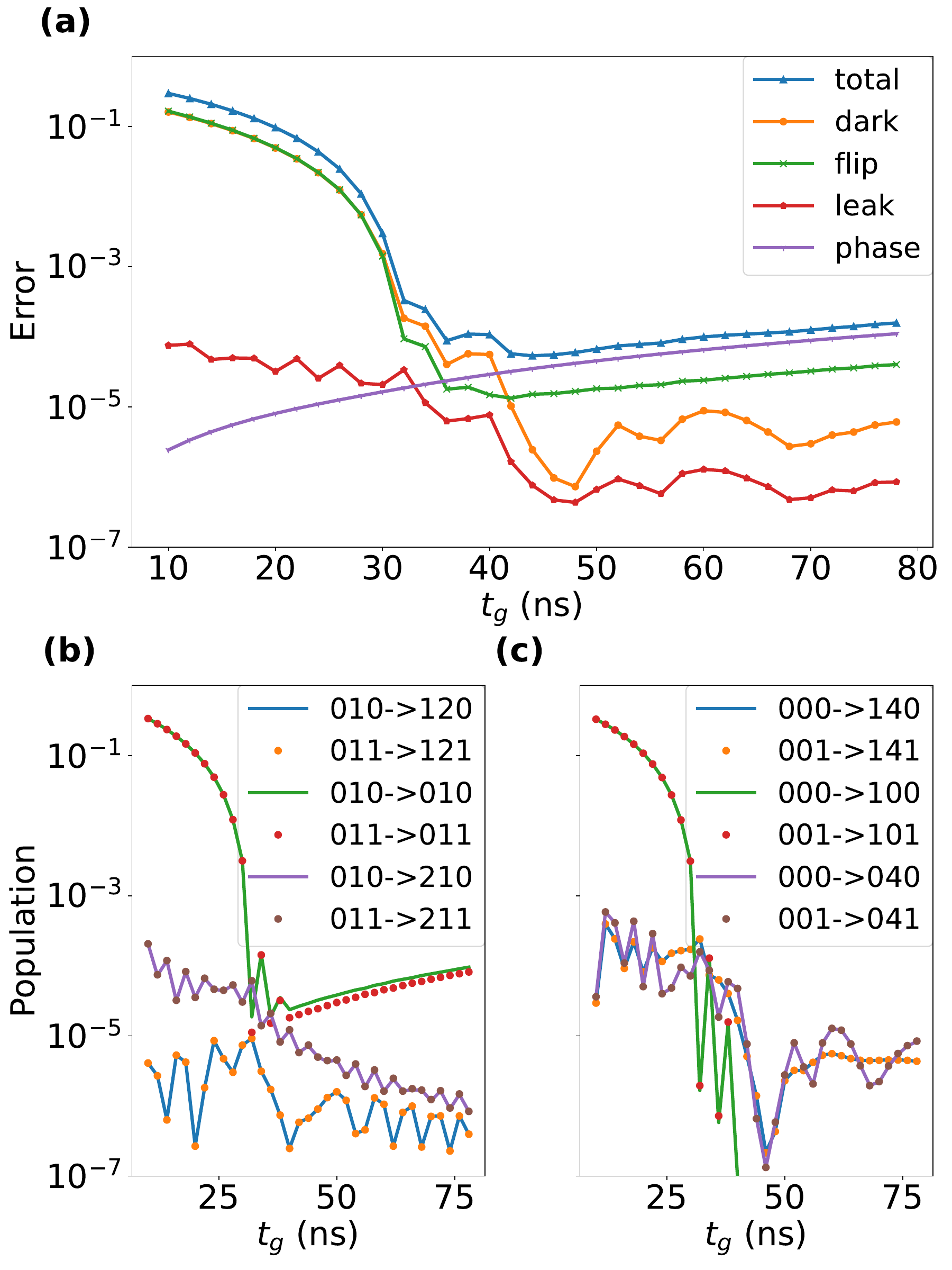}
\caption{(a): Categorized errors in optimized $CX_1$ gates as a function of $t_g$ in the 3-qubit system. Dominant (b) bright state and (c) dark state error populations using the same gates optimized for minimal overall error. As expected, ``phase'' errors from spectator-induced variation in $ZZ$ accumulation between the active transmon-fluxonium pair set the lower bound for gate performance. This trend's smooth and predictable behavior enables efficient parameter search using our two-step optimization process. Specifically, we first calibrate population transfers, then correct phases.}\label{fig:fig6}
\end{figure}
To explore the system and drive parameter landscapes, we consider different values of $J$, and $t_g$. For each set of different system parameters, we re-optimize drive parameters for maximum fidelity. The high $J$ and high $t_g$ regimes show a monotonic increase of $\mathcal{E}$ caused by an unavoidable detuning of the optimal drive from the transmon frequency. This detuning arises from the spectator qubit's state shifting the transmon's dressed $\ket0 \leftrightarrow \ket1$ transition frequency as shown in Fig.~\ref{fig:fig5}. The resulting parasitic effect can be analyzed as a consequence of transmon-fluxonium $ZZ$, which is written as $f_{ZZ} = f_{10-11} - f_{00-01}$ in either FT subspace. We choose the spectator qubit's FT subspace, such that $f_{ZZ}$ is precisely the difference in transmon transition frequency conditioned on the spectator state. For the FTF system, we can use $f_{ZZ}/2$ to estimate the minimum unavoidable detuning of a Rabi drive of the transmon from bright initial states involving the active fluxonium. The selective darkening approach allows for well behaved and frequency-insensitive dark states by choosing the correct ratio of drive amplitudes. Due to this dark-state transition stability, errors associated with bright-state transitions dominate in the regime of high gate time. The dependence on detuning of two bright-state transmon Rabi flip populations dressed by different spectator states is plotted in Fig.~\ref{fig:fig5}(e). To demonstrate the correspondence of the spectator-biased imperfect transmon rotations with overall performance, we also plot gate fidelity as a function of drive detuning without re-optimization at every step. By placing the spectator in either state $\ket0$ or $\ket1$, and sweeping error for the targeted bright transition, we observe the extrema corresponding to the optimal drive frequency for that initial state. By comparing the gate's optimal drive frequency to points in each of the two spectator-biased sweeps, we see that it stays between the ideal frequencies corresponding to having the spectator in state $\ket0$ and state $\ket1$. Note that the two sweep frequencies correspond to each computational spectator basis state after accounting for drive-induced Stark shifts. The effect of Stark shifts grows with drive strength, and therefore decreases with higher $J$ for our re-optimized gates. This relationship is also seen in Fig.~\ref{fig:fig5}, where the detuning of $f_d$ from the non-driven bright-state frequencies decreases when increasing $J$. However, the stark shift itself is quite insensitive to the spectator state, so its strength is not nearly as important as the increased spectator-dependent frequency splitting at high $J$. For each value of $J$, the optimal pulse frequency point maintains roughly equal detuning from each bright transition frequency. In the presence of a drive, this unavoidable detuning generates ``flip'' and ``phase'' errors, which dominate our error budget at gate times above 44 ns.

\begin{figure}
\includegraphics[width=0.475\textwidth]{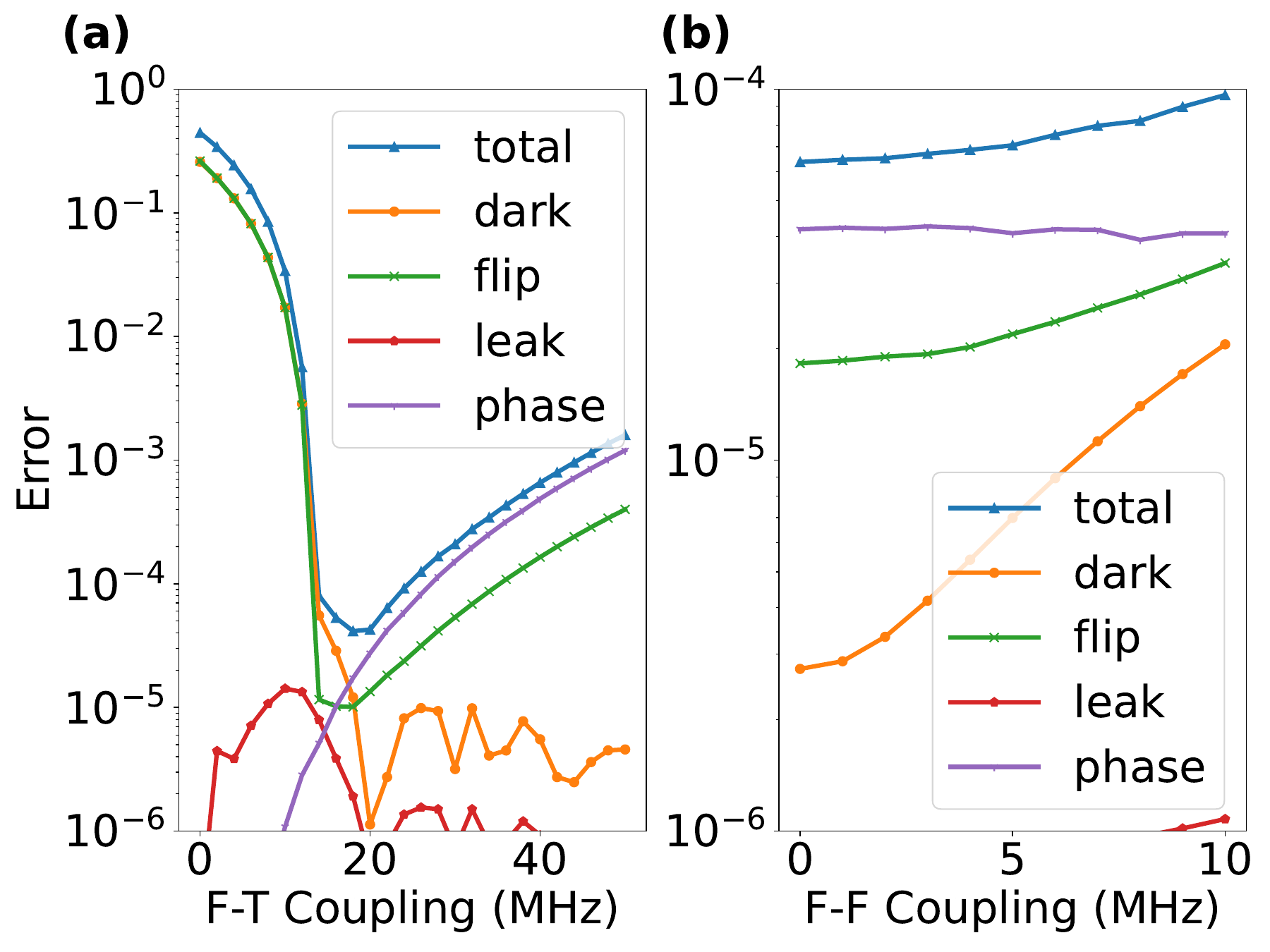}
\caption{Categorized gate error as a function of (a) $J/h$ and (b) $I/h$ for optimized $CX_1$ using $t_g$ = 50 ns. Phase error dominates in stable parameter regimes when $J \gg I$. This error is the result of $ZZ$-type transmon frequency variance based on the spectator state.}\label{fig:fig7}
\end{figure}

Since the drive's Stark effect shifts the two bright-transition frequencies with less than $1\%$ difference in magnitude when $J/h = 22$ MHz, the effective detuning of $CX_1$ Rabi oscillations is estimated well by $\Delta_{\rm eff} \approx \left|f_{011-111} - f_{010-110}\right|/2$. An individual ``flip'' error transition then has an associated probability of approximately $P_{\rm err} \approx (\Delta_{\rm eff}/\Omega_{\rm Rabi})^2 = 4 \Delta_{\rm eff}^2 t_g^2$ during the gate. Clearly, in addition to ``phase'' errors, detuning in the Rabi drive also causes the accumulation of more ``flip'' errors at longer gate times. This can be seen through the increase in $\ket{011} \to \ket{011}$ probability during the gate when going from a 35 ns pulse to a 100 ns pulse in Fig~\ref{fig:fig5}. As such, optimizing gate time and fluxonium-transmon coupling strength corresponds to balancing detuning-induced ``phase'' and ``flip'' errors at long gate times with counter-rotating dynamics in short gates. We find that the $CX_1$ gate exhibits a valley of low $\mathcal{E}$ in Fig.~\ref{fig:fig6} and Fig.~\ref{fig:fig7} by tuning $t_g$ and $J/h$ to about $44$, ns and $20$ MHz, respectively.

On the other hand, fluxonium-fluxonium coupling must be kept arbitrarily weak for optimal performance. Increasing its strenth hybridizes fluxonium energy levels, introducing a crosstalk channel for the strongly driven control qubit. As shown in Fig.~\ref{fig:fig7}, the presence of nonzero $I$, and subsequent involvement of the spectator qubit ruins the quality of selectively darkened transitions and bright transitions. Thus, both spectator-control and spectator-target coupling terms introduce parasitic effects to the gate through different mechanisms. In this sense, the spectator fluxonium effectively introduces environmental dephasing into the active qubit subspace.

\section{Applications of fluxonium-transmon \textsc{cnot} gates}
\label{sec:IV}
\begin{figure*}[ht]
    \centering
    \includegraphics[width=1\textwidth]{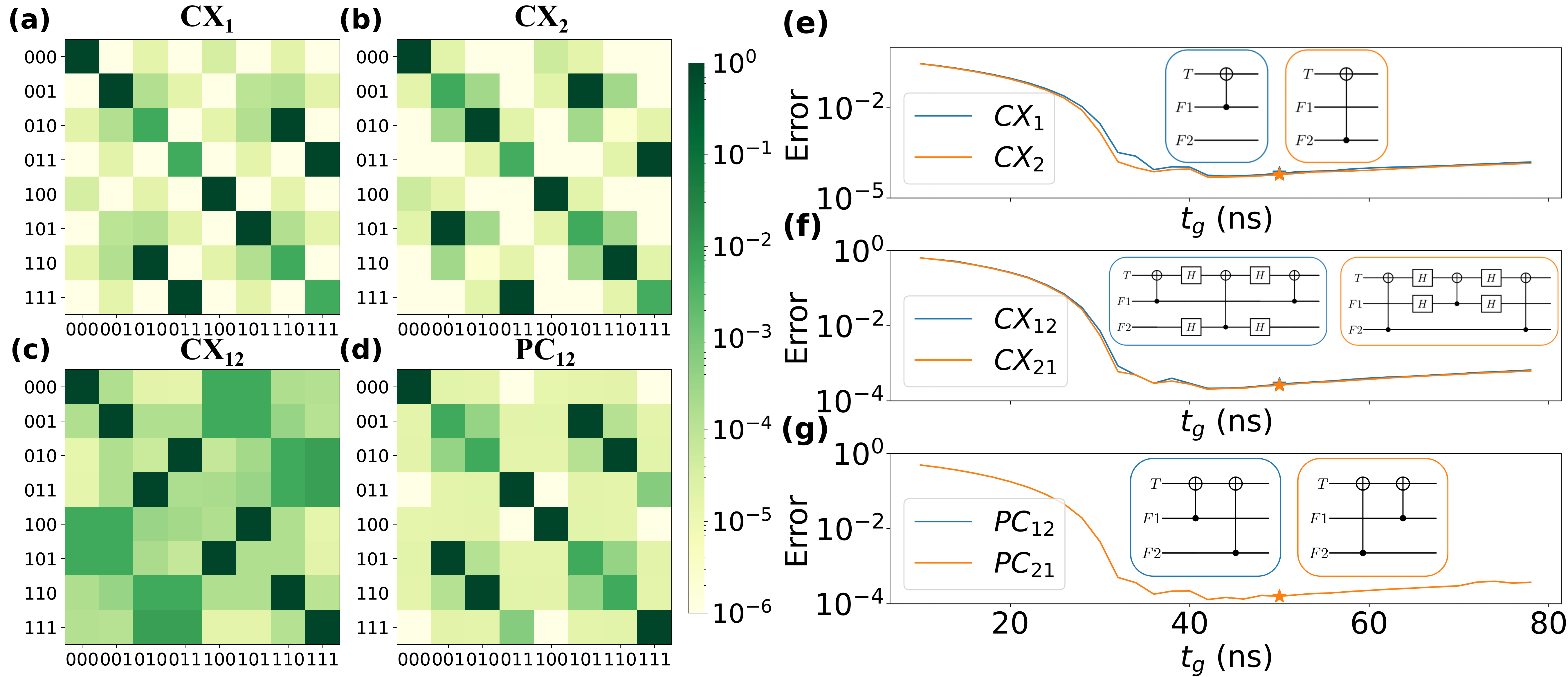}
    \caption{Evolution operator diagrams for optimized (a) $CX_1$, (b) $CX_2$, (c) $CX_{12}$, and (d) $PC_{12}$ gates with $t_g$ = 50 ns individual CR pulse time. Comparisons of (e) $CX_1$ and $CX_2$, (f) logical fluxonium-fluxonium \textsc{cnot} gates $CX_{12}$ and $CX_{21}$, and (g) sequential parity checks $PC_{12}$ and $PC_{21}$ as functions of total gate time are also shown. For parity check errors, we use a comparison of the full $8\times8$ operator with the product of ideal \textsc{cnot} gates. Compound fluxonium-fluxonium gate errors are extracted from the effective evolution operator defined by Eq.~\eqref{eq:compoundgate0s}. To isolate the effects of $CX_1$ and $CX_2$ gate errors, we assume that single qubit rotations take no time and have perfect fidelity. Note that for the compound gates, the full operation takes a total time $3t_g$, and for parity checks, it takes $2t_g$.}
    \label{fig:fig8}
\end{figure*}

The entangling gates demonstrated in previous sections enable various operations, targeting utility in large-scale quantum computers. For example, a $CX_{\alpha}$ gate can be used to map the state of F$\alpha$ to the transmon. When followed by readout of the transmon, F$\alpha$ is effectively read without demolition. Additionally, sequences of $CX_1$ and $CX_2$ gates can be used for fluxonium-fluxonium parity checks and gates. The performance of these applications in our full FTF system using 50 ns CR pulses is shown in Fig.~\ref{fig:fig8}. With access to such gates, logical readout and syndrome extraction in a surface code setting would only require resonators coupled to the transmons.

\subsection{Fluxonium readout}
\subsubsection{FT system}

In a setup where qubit T does not store computational data, it can be kept in the ground state and used for ancillary operations, including neighboring fluxonium readout. We propose doing so using our previously described CR \textsc{cnot} gates between the fluxonium of interest and the transmon. For an ideal \textsc{cnot} gate, we expect the following evolution:
\begin{equation}
\label{eq:CXreadout}
(c_0\ket{0_F}+c_1\ket{1_F})\otimes \ket{0_T} 
\xrightarrow{\rm CNOT}
c_0\ket{0_F,0_T}+e^{i\theta}c_1\ket{1_F,1_T}.
\end{equation}
After this operation, transmon readout can be performed. One benefit of this approach is the experimentally demonstrated speed, fidelity, and robustness of transmon readout~\cite{shillito2022dynamics, cohen2023reminiscence, chen2023transmon, swiadek2024enhancing, kurilovich2025high, spring2025fast}.
We first discuss the effect of fluxonium-state assignment to the transmon in the FT system. We use the optimized \textsc{cnot} gate discussed in Sec.~\ref{sec:2qubit}, and apply it to the initial state in Eq.~\eqref{eq:CXreadout}. The final state of the system is a superposition of correct states with states that cause incorrect readout and/or demolition of the data qubit state. We define non-demolition assignment fidelity for this process as $\mathcal{F}_{\rm QND} = (P_{00\to 00} + P_{10\to 11})/2$ where
\begin{equation}
    P_{ij\to i'j'} = \left|\bra{i'j'}\hat{U}_{\rm CNOT}\ket{ij}\right|^2.
\end{equation}
If readout is performed at the end of a circuit, incidental fluxonium state demolition incurs no negative effects. For such situations, we define demolition assignment fidelity as $\mathcal{F}_{\rm nonQND} = \mathcal{F}_{\rm QND} + (P_{00\to 10} + P_{10\to 01})/2$. The correction terms associated with demolition assignment involve a flip of the control qubit. Because of our gigahertz-order detuning between the control and target qubit, such transitions stay below $10^{-9}$ probability for gate times higher than 40 ns. With this subtlety proving negligible in practice, essentially all errors present in the gate affect both QND and non-QND assignment in the context of fluxonium readout.

\subsubsection{FTF system}
The same approach can be extended to the FTF system. We now index states as $\ket{i_T j_{\rm{F} \alpha} k_{\rm{F} \beta}}$ and examine the use of $CX_{\alpha}$ to perform readout of F$\alpha$. A $CX_{\alpha}$ gate has the ideal evolution
\begin{equation}
\ket{0} \otimes (c_0\ket{0}+c_1\ket{1}) \otimes \ket{\beta}
\xrightarrow{CX_{\alpha}}
c_0\ket{0,0,\beta}+e^{i\theta}c_1\ket{1,1,\beta}.
\end{equation}
Depending on the conditions for readout to qualify as successful, different probability amplitudes can be considered. Now using shorthand notation
\begin{equation}
    P_{ijk\to i'j'k'} = \left|\bra{i'j'k'}\hat{U}_{CX_{\alpha}}\ket{ijk}\right|^2,
\end{equation}
the probability of correctly preparing the transmon for non-demolition readout~\cite{ralph2006quantum} is $\mathcal{F}_{\rm QND} = (P_{00\beta \to 00\beta} + P_{01\beta \to 11\beta})/2$. We also define demolition errors such that they do not disturb immediate readout, $\ket{\beta}$ state preservation, or unitarity. Errors where the control qubit experiences an undesired flip are the only ones following such demolition-readout criteria. The combined effect of such errors is described by $\mathcal{F}_{\rm nonQND} = \mathcal{F}_{\rm QND} + (P_{00\beta \to 01\beta} + P_{01\beta \to 10\beta})/2$. Due to negligible effects of $CX_\alpha$ on the control qubit, for all gate time values tested, $\mathcal{F}_{\rm nonQND} - \mathcal{F}_{\rm QND} < 10^{-8}$. In a given context of readout correctness criteria, it is useful to define $\mathcal{E}_{\rm read} = 1 - \sum_iP_i$ where $\{P_i\}$ is the set of all acceptable transitions. As demonstrated in Fig.~\ref{fig:fig6}, dominant $CX_\alpha$ errors in this context result in leakage or incorrect readout. This term quantifies the additional assignment error associated with measuring F$\alpha$ through T. It is important to note that $\mathcal{E}_{\rm read}$ does not account for infidelity associated with native transmon readout, but instead provides an upper bound for neighboring-fluxonium readout fidelity.
\subsection{Parity checks of two fluxoniums}
\begin{figure}
\includegraphics[width=0.475\textwidth]{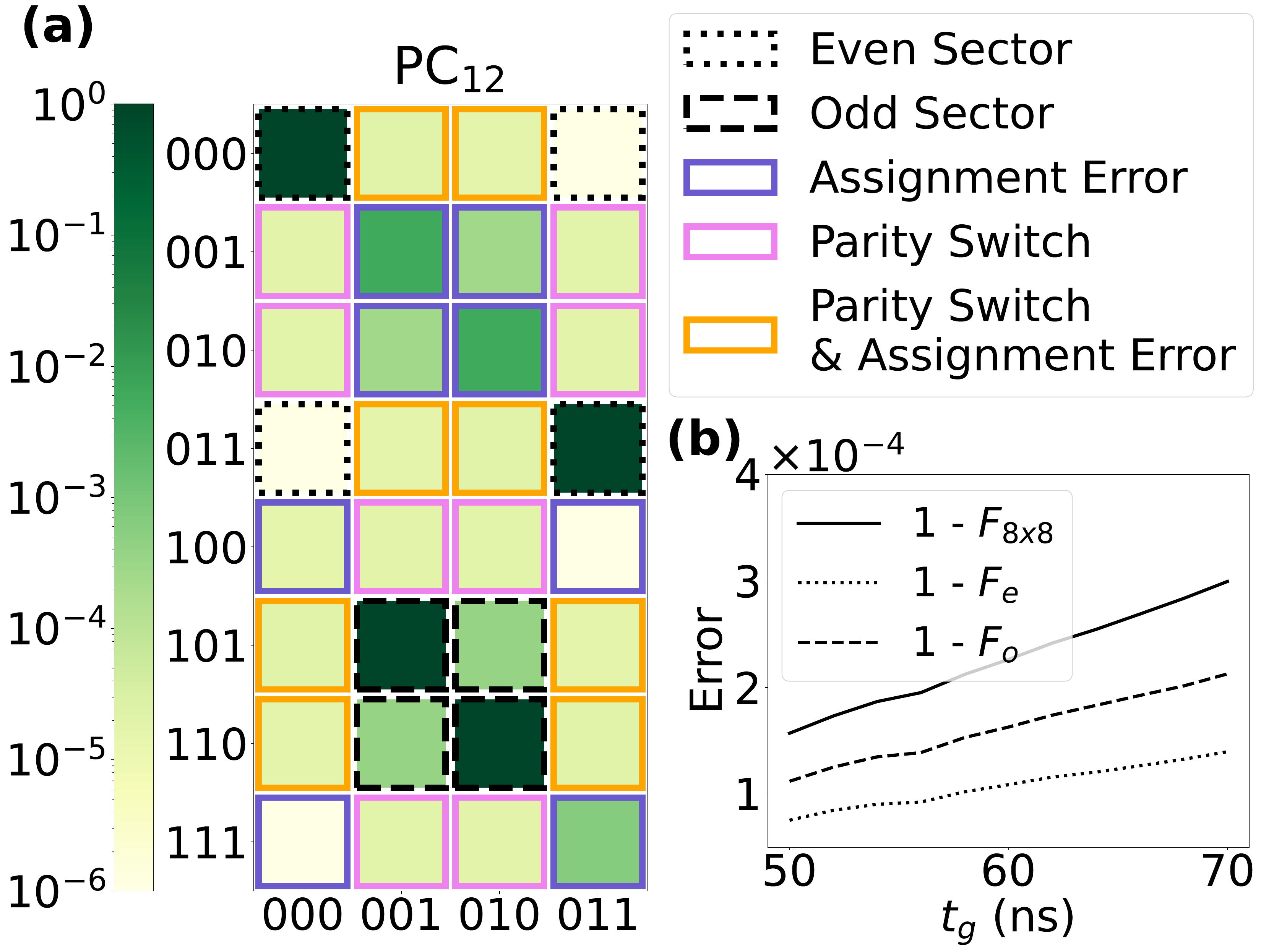}
\caption{(a): Grouped $PC_{12}$ matrix elements using optimized 50 ns $CX_1$ and $CX_2$ gates. (b): Error for the same gate within different subspaces as a function of single CR pulse time in the stable high-time regime. Matrix elements are primarily separated following the even and odd parity matrices defined in Eq.~\eqref{eq:evenpcgate}, and Eq.~\eqref{eq:oddpcgate} respectively. Remaining elements are error transitions which we aggregate based on their retention of fluxonium parity, and assignment accuracy to the transmon. Error within these even and odd parity subspaces is plotted, as well as full $8\times8$ error using a sequence of perfect $CX_1$ and $CX_2$ operations as the ideal gate.}\label{fig:fig9}
\end{figure}
Another proposed use case for $CX_{\alpha}$ gates is to perform parity checks between F1 and F2 that are measured via the transmon. Following the strategy of keeping T in state $\ket{0}$ before the operation, we can construct the parity check process as the product of $CX_1$ and $CX_2$ gates: $PC_{\alpha \beta} = CX_\beta \cdot CX_\alpha$. We continue indexing as $\ket{i_T j_{\rm{F} \alpha} k_{\rm{F} \beta}}$ for generality unless otherwise stated. Using ideal $CX_\alpha$ and $CX_\beta$ gates, evolution of the parity check sequence follows
\begin{multline}
\label{eq:parity_check}
    [c_{00}\ket{000} + c_{11}\ket{011}] + [c_{10}\ket{010} + c_{01}\ket{001}] \xrightarrow{PC} \\
    [e^{i\theta_{00}}c_{00}\ket{000} + e^{i\theta_{11}}c_{11}\ket{011}] + [e^{i\theta_{10}}c_{10}\ket{110} + e^{i\theta_{01}}c_{01}\ket{101}],
\end{multline}
with $\theta_{ij}=0$.
While preserving coherence within even or odd sectors, we can relax the phase equality requirement to less stringent conditions $\theta_{00}=\theta_{11}$, and $\theta_{01}=\theta_{10}$. But even the latter conditions can be lifted by applying additional $R_Z$ gates to correct the differences $\theta_{00}-\theta_{11}$ and $\theta_{01}-\theta_{10}$ conditioned on the result of parity readout. For numerical simulations, we use $CX_\alpha$ gates that have already been phase corrected using the previously discussed two-step optimization. The parity check evolution operator in the computation subspace is characterized by the matrix elements $\bra{i'j'k'} U_{\rm PC} \ket{ijk}$. To establish informative parity check fidelity metrics, we separate the dynamics of even and odd initial fluxonium states. For the even fluxonium configuration, we can define a $2\times 2$ matrix that does not change the transmon state or the fluxonium parity as
\begin{equation}
\label{eq:evenpcgate}
\hat U_{e} = 
\left(
\begin{array}{cc}
\bra{000}U_{\rm PC}\ket{000} & \bra{000}U_{\rm PC}\ket{011} \\
\bra{011}U_{\rm PC}\ket{000} & \bra{011}U_{\rm PC}\ket{011}
\end{array}
\right).
\end{equation}
For odd fluxonium configurations that ideally flip the transmon from $\ket{0_T}$ to $\ket{1_T}$, we have 
\begin{equation}
\hat U_{o} = 
\left(
\begin{array}{cc}
\bra{101}U_{\rm PC}\ket{001} & \bra{101}U_{\rm PC}\ket{010} \\
\bra{110}U_{\rm PC}\ket{001} & \bra{110}U_{\rm PC}\ket{010}
\label{eq:oddpcgate}
\end{array}
\right).
\end{equation}
The ideal parity check in both the even and odd subspaces is the $2\times2$ identity matrix. Thus, the fidelity of the parity check within each subspace can be expressed in terms of the above matrices as
\begin{equation}
\label{eq:paritycheckevenoddfid}
    \mathcal{F}_p =\frac{\mathrm{Tr} \{\hat U_p \hat U_p^\dag\} +|{\mathrm{Tr}}\{\hat U_p\}|^2}{6},\quad p\in\{e,o\}.
\end{equation}
Note that off-diagonal elements of $\hat U_p$ describe the error introduced by the parity check on the fluxonium states without disturbing parity or assignment. Parity switching and transmon flips act the same as leakage by reducing the norm of the considered matrix. Our parity separation approach accounts for 8 of the $8\times4$ matrix elements describing computational parity check evolution while requiring the transmon to have the initial state $\ket0$. The remaining elements can be grouped into additional matrices. A $4\times2$ assignment error matrix can be defined where parity is conserved but the transmon state is demolished. A parity switch matrix where the transmon shows original parity also corresponds to a $4\times2$ matrix. Additionally, parity switches where the transmon also indicates the wrong initial parity represent the final $4\times2$ matrix necessary for a complete description. This grouping of matrix elements using data from a simulated $PC_{12}$ gate is demonstrated in Fig.~\ref{fig:fig9}. To compare the even and odd fidelities defined in Eq.~\eqref{eq:paritycheckevenoddfid}, we also evaluate and plot gate error of the full parity check operator in Fig.~\ref{fig:fig9}. For this, we use Eq.~\eqref{eq:fidelity} to compute $\mathcal{F}_{8\times8}$ as the fidelity for the $8\times 8$ product of $CX$ evolution operators compared to the product of corresponding ideal gates. All of our parity check error definitions yield gates with $\mathcal{F} > 99.98\%$ for pulse times between 42 ns and 56 ns. In the same regime, they also stay within a range of $1.03 \times 10^{-4}$, and we consistently see $\mathcal{F}_{e} > \mathcal{F}_{o} > \mathcal{F}_{8\times8}$. This is consistent with our previous observations of dark initial states not dominating the individual $CX_{\alpha}$ error budgets.
\subsection{Compound \textsc{cnot} gate between two fluxoniums}
\label{sec:compound}
\subsubsection{Process fidelity approach}
\begin{figure}
\includegraphics[width=0.375\textwidth]{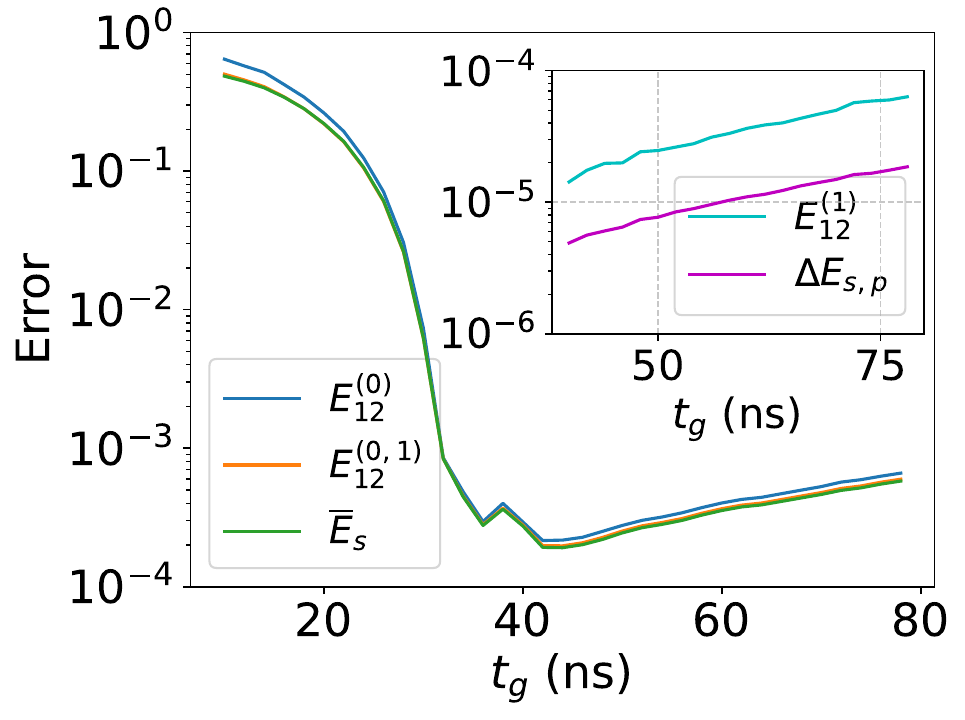}
\caption{Compound fluxonium-fluxonium \textsc{cnot} gate error with control qubit F1 and target qubit F2 as a function of single CR pulse time using all three fidelity metrics defined in Sec.~\ref{sec:compound}. The inset plot shows error differences between each nearest error trend. Namely, $E_{12}^{(1)} = E_{12}^{(0)} - E_{12}^{(0,1)}$ acts as a correction term for allowing the transmon to reach final state $\ket{1}$ using our process fidelity approach, and $\Delta E_{s,p} = E_{12}^{(0,1)} - \bar{E}_s$ compares the two fidelity measurement approaches.}\label{fig:fig10}
\end{figure}

With $CX_1$, $CX_2$, and local Hadamards, a \textsc{cnot} operation between F$\alpha$ (control) and F$\beta$ (target) can also be performed. When T is initialized to $\ket{0}$, and both fluxoniums are in arbitrary computational states, the sequence $CX_{\alpha \beta}$ for a logical \textsc{cnot} gate between F$\alpha$ (control) and F$\beta$ (target) is
\begin{equation}
   \hat{U}_{CX_{\alpha \beta}} = \hat{U}_{CX_{\alpha}} (\hat{H} \otimes \hat{I} \otimes \hat{H}) \hat{U}_{CX_{\beta}} (\hat{H} \otimes \hat{I} \otimes \hat{H}) \hat{U}_{CX_{\alpha}}.
\end{equation}
Here, ``$\otimes$'' represents a Kronecker product of matrices which are indexed in the order of T, F$\alpha$, F$\beta$. Using individually optimized $CX_1$ and $CX_2$ gates, now alongside perfect virtual Hadamard gates, we can construct a sequence for $\hat{U}_{FF}^{\alpha}$ where all error originates from the fluxonium-transmon \textsc{cnot} gates. When determining the overall fidelity of $\hat{U}_{FF}^{\alpha}$, we only want to consider valid initial states. Because T is in state $\ket{0}$ before and ideally after the sequence, matrix elements for the process conserving the transmon state are
\begin{equation}
\label{eq:compoundgate0s}
    \hat{U}_{ij,kl}^{(0\to 0)} = \bra{0,i,j}\hat{U}_{CX_{\alpha \beta}}\ket{0,k,l}.
\end{equation}
We may also consider the case when a \textsc{cnot} gate between fluxoniums occurs, but the final state of T flips to $\ket1$ to be successful. For this scenario, a reset of T back to $\ket0$ is necessary before successive operations with the transmon. The effective gate encompassing transmon flip contributions has matrix elements
\begin{equation}
    \hat{U}_{ij,kl}^{(0\to 1)} = \bra{1,i,j}\hat{U}_{CX_{\alpha \beta}}\ket{0,k,l}.
\end{equation}
Since the effective evolution operator is in the fluxonium-fluxonium subspace, we calculate effective fidelity as
\begin{equation}
    \mathcal{F}_{\alpha \beta}^{(i)} = \frac{\mathrm{Tr}(\hat{U}_i^{\dag}\hat{U}_i) + \left|\mathrm{Tr}(\hat{U}_{\rm CX}^{\dag}\hat{U}_i)\right|^2}{20}, \quad U_i = U^{(0\to i)}.
\end{equation}
Then, gate error is $E^{(0)}_{\alpha \beta} = 1 - \mathcal{F}_{\alpha \beta}^{(0)}$ when only accepting $\ket0$ as the final transmon state, and $E^{(0, 1)}_{\alpha \beta} = 1 - \mathcal{F}_{\alpha \beta}^{(0)} - \mathcal{F}_{\alpha \beta}^{(1)}$ when accepting any computational final transmon state. It follows that the correction term associated with accepting the $\ket{1}$ final transmon state is $E_{\alpha \beta}^{(1)} = E_{\alpha \beta}^{(0)} - E_{\alpha \beta}^{(0,1)}$. Using optimized $CX_1$ and $CX_2$ from Sec.~\ref{sec:2qubit}, we construct logical fluxonium-fluxonium gates with effective fidelity $\mathcal{F}_{\rm eff}>99.97\%$. The correction associated with allowing transmon flips is on the order of $10^{-5}$ in the sufficiently high gate time regime, as seen in Fig.~\ref{fig:fig10}.
\subsubsection{State fidelity approach}
An alternative approach to characterizing our fluxonium-fluxonium gate is through comparison of simulated and expected final qubit states. As before, valid initial states must have the transmon set to $\ket0$. For uniform sampling of this space, we define a set of initial states $\mathcal{S}$ to contain all combinations $\ket{0_T}\otimes\ket{j}\otimes\ket{k}$ where $j,k \in \{\ket{0},\ket{1}, \ket{+x}, \ket{-x}, \ket{+y}, \ket{-y}\}$. For each initial state $\ket{s} \in \mathcal{S}$, we can construct the density matrix of its final state, and trace out the transmon's degree of freedom as
\begin{equation}
\rho_s = \sum_{j,j',k,k'=0}^1\ket{j,k}\bra{j',k'}\left[\sum_{i=0}^1 \bra{i,j,k}U\ket{s}\bra{s}U^{\dagger}\ket{i,j',k'}\right].
\end{equation}
We then evaluate fidelity using the state fidelity expression
\begin{equation}
    \mathcal{F}_s(\rho_{s}, \sigma_s) = \left( \mathrm{Tr} \left[ \sqrt{ \sqrt{\rho_{s}} \sigma_s \sqrt{\rho_{s}} } \right] \right)^2.
\end{equation}
Here, $\sigma_s$ is the ideal final density matrix of the two-fluxonium subspace starting from the initial state $\ket{s}$. For all tested values of $t_g$, the average final state error follows $\bar{E}_s = \sum_{s\in \mathcal{S}}[{1 - \mathcal{F}_s(\rho_{s}, \sigma_s)}]/\left| \mathcal{S} \right| < E_{12}^{(0, 1)} < E_{12}^{(0)}$, as seen in Fig.~\ref{fig:fig10}. To quantify the difference in error for the state fidelity and process fidelity approaches, we also plot $\Delta E_{s,p} = E_{12}^{(0,1)} - \bar{E}_s$.
\begin{figure}
\includegraphics[width=0.375\textwidth]{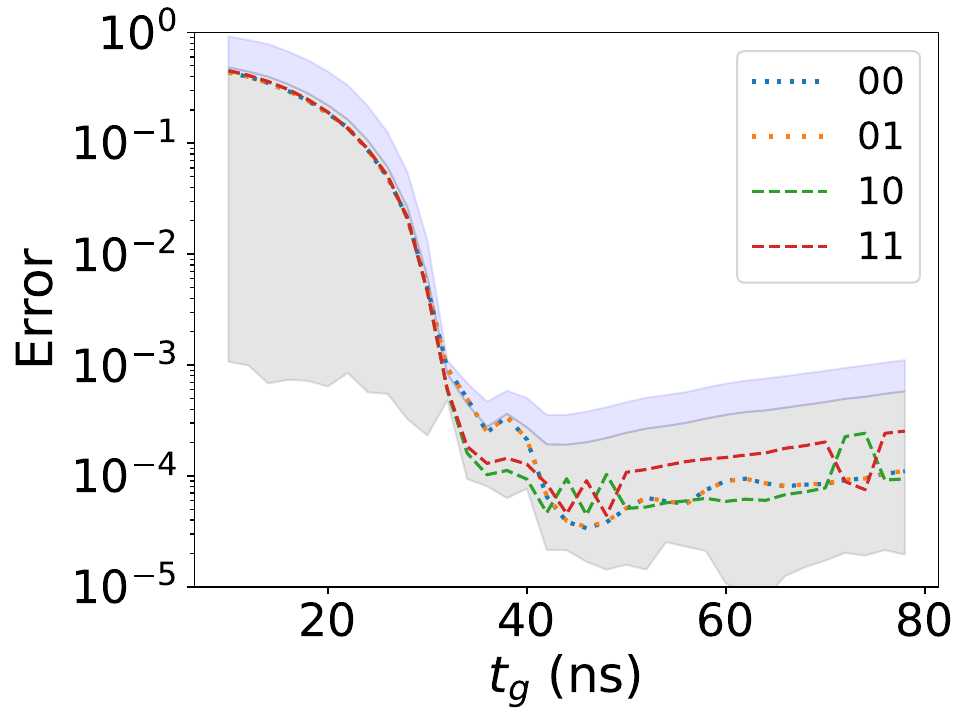}
\caption{Compound fluxonium-fluxonium \textsc{cnot} final state errors as a function of single CR pulse time using the state fidelity approach. Trends labeled as $ij$ correspond to initial states $\ket{0ij}$. The difference in shading on the plot shows the average final state fidelity of all fluxonium basis state combinations from the set $\mathcal{S}_F = \{\ket{0},\ket{1}, \ket{+x}, \ket{-x}, \ket{+y}, \ket{-y}\}$. The upper and lower shading boundaries correspond to the highest and lowest errors, respectively, from any initial state $s \in \mathcal{S}$. Jumping behavior of individual trends in the stable high-time regime is a result of re-optimization at each step. The optimization is for individual F-T gate fidelity, and thus, interference of several gates in the sequence can result in different contributions with similar magnitudes at each point.}\label{fig:fig11}
\end{figure}
\section{Discussion and conclusions}
\label{sec:V}

In this work, we have analyzed the coherent performance of cross-resonance \textsc{cnot} gates in a dual-species fluxonium-transmon systems. The foundation of this analysis is a single \textsc{cnot} gate between a fluxonium and a transmon with a high fidelity of over $99.994\%$. Building on this, we show that a sequence of these gates can be used to perform a two-fluxonium parity check gate with an effective fidelity exceeding $99.98\%$. We also designed a transmon-facilitated \textsc{cnot} gate between the two fluxoniums, achieving a fidelity greater than $99.97\%$. A key advantage of these gates is their simplicity, relying on only a charge drive with a basic pulse shape. Additionally, the gate scheme is robust to variation in qubit Josephson energies, as demonstrated in Appendix~\ref{appendix:errorbudget}. This makes them excellent candidates for near-term experimental implementation.

The weak qubit-qubit coupling in our design is crucial, as it allows for scalability to larger quantum processors, where localized gate operations can be performed on smaller sub-sections of the system without significant crosstalk. The primary error limiting the fidelity of our $CX_{\alpha}$ gates is rooted in spectator-induced transmon frequency shifts. This issue, arising from the presence of other qubits, detunes the drive from its ideal frequency. We maintain low coupling between fluxoniums and neighboring transmons to mitigate this effect. The operation of this dual-species system does not require direct coupling between fluxoniums, thus reducing their $ZZ$ interaction. This direct coupling would otherwise enhance the variance in strength of $ZZ$ terms in systems with many fluxoniums, resulting in an intrinsic limit on gate performance. To completely remove the weak indirect $ZZ$ coupling, a promising approach would be to use AC-Stark shift-facilitated $ZZ$ drives~\cite{xiong2022arbitrary}. This would further enhance the robustness and fidelity of gates in our dual-species fluxonium-transmon approach for developing scalable quantum processors.

We also note that in a scaled grid of our proposed architecture, the impact of a transmon spectator in a TFT-type subsystem may be considered. For our cross-resonant gate, the transmon-spectator induced error is then determined by the proximity of target and spectator transmon frequencies. Resonance of the two qubits is problematic, since the shared fluxonium neighbor is driven close to their frequency and introduces a channel for crosstalk through hybridization of their eigenstates. As a result, the fluxonium-coupled drive generates an effective drive of the spectator with strength inversely proportional to their detuning $\Delta_T$. As such, we can estimate the error induced by our transmon spectator as $P_{\rm err} \sim (\Delta_T t_g)^{-2}$.

Our proposed gate times are short in comparison to modern fluxonium~\cite{somoroff2023millisecond} and transmon~\cite{bland20252d} coherence times, and thus we do not explicitly consider incoherent dynamics. Even so, these effects are easily estimated with additional terms of the form $\mathcal{E}_{\rm de}^{(i)} \approx 1 - \exp[-t_g (1/T_1^{(i)} + 1/T_{\phi}^{(i)})]$, for qubit $i$ with known relaxation time $T_1^{^{(i)}}$, dephasing time $T_{\phi}^{^{(i)}}$, and gate time $t_g$~\cite{ziman2005all, kofman2009two}. 

\begin{acknowledgments}
We would like to thank Jiakai Wang, Tanvir Ahmed Masum, and Konstantin Nesterov for insightful discussions. Some parts of our numerical simulations were performed using the QuTiP~\cite{Johansson2012, Johansson2013}, and scqubits~\cite{groszkowski2021scqubits} python packages. This research was supported by the ARO GASP (contract No. W911-NF23-10093) program.
\end{acknowledgments}
\appendix

{

\section{Ideal Qubit Model}
\label{app:ideal_qubit_cr}

For completeness, we provide a detailed derivation of the effective cross-resonance (CR) Hamiltonian used in Sec.~\ref{sec:III}. In particular, we explicitly construct the rotating-frame transformation, the time-dependent Schrieffer--Wolff (SW) generator, and estimate the magnitude of neglected terms.


We consider two coupled qubits, labeled $A$ (control) and $B$ (target), with a transverse interaction and a drive applied to qubit $A$ at frequency $\omega_B$:
\begin{equation}
\hat{H}(t) = \frac{\omega_A}{2} \sigma_z^A + \frac{\omega_B}{2} \sigma_z^B 
+ J \sigma_x^A \sigma_x^B 
+ \hat{H}_{d}(t),
\end{equation}
where the drive term for the cross-resonant contribution has the form:
\begin{equation}
    \hat{H}_{d}(t) = \Omega_d \cos(\omega_B t)\sigma_x^A
\end{equation}

We split the Hamiltonian as
\begin{equation}
\hat{H}(t) = \hat{H}_0  + \hat{V}(t) ,
\quad
\hat{H}_0 = \frac{\omega_A}{2} \sigma_z^A + \frac{\omega_B}{2} \sigma_z^B
\end{equation}
and use the interaction representation defined by $\hat{H}_0$:
\begin{equation}
\hat{U}_0(t) = e^{-i \hat{H}_0 t}, \quad
\tilde{H}(t) = \hat{U}_0^\dagger(t) \hat{V}(t) \hat{U}_0(t).
\end{equation}
Using
\begin{equation}
\sigma_x = \sigma_+ + \sigma_-, \quad
\sigma_\pm(t) = \sigma_\pm e^{\pm i \omega t},
\end{equation}
we write 
the coupling term as
\begin{align}
\nonumber
\sigma_x^A \sigma_x^B &= (\sigma_+^A + \sigma_-^A)(\sigma_+^B + \sigma_-^B) \\
&= \sigma_+^A \sigma_-^B + \sigma_-^A \sigma_+^B + \text{fast terms}.
\end{align}
After transformation and applying the rotating-wave approximation (RWA), we retain only near-resonant terms:
\begin{equation}
\tilde{H}_J(t) =
J \left(
\sigma_+^A \sigma_-^B e^{i\Delta t}
+ \sigma_-^A \sigma_+^B e^{-i\Delta t}
\right),
\end{equation}
where $\Delta = \omega_A - \omega_B$.

Similarly, the drive term in the RWA becomes

\begin{equation}
\tilde{H}_d(t) =
\frac{\Omega_d}{2}
\left(
\sigma_+^A e^{i\Delta t}
+ \sigma_-^A e^{-i\Delta t}
\right).
\end{equation}

Combining the above expressions, we obtain 
\begin{equation}
\hat{H}_{\mathrm{rot}}(t) =
\hat{H}_{+} e^{i\Delta t} + \hat{H}_{-} e^{-i\Delta t},
\end{equation}
with
\begin{equation}
\hat{H}_{+} = J \sigma_+^A \sigma_-^B + \frac{\Omega_d}{2}\sigma_+^A,
\quad
\hat{H}_{-} = J \sigma_-^A \sigma_+^B + \frac{\Omega_d}{2}\sigma_-^A.
\end{equation}

We now construct  the time-dependent Schrieffer--Wolff transformation by introducing 
a unitary transformation
\begin{equation}
\hat{U}(t) = e^{\hat{S}(t)}, \quad \hat{S}^\dagger = -\hat{S},
\end{equation}
such that the transformed Hamiltonian
\begin{equation}
\hat{H}_{\mathrm{eff}} = e^{\hat{S}} \hat{H}_{\mathrm{rot}} e^{-\hat{S}} - i e^{\hat{S}} \partial_t e^{-\hat{S}}
\end{equation}
becomes time-independent to leading order.
Expanding to second order:
\begin{equation}
\hat{H}_{\mathrm{eff}} =
\hat{H}_{\mathrm{rot}} + [\hat{S}, \hat{H}_{\mathrm{rot}}]
- i \partial_t \hat{S}
+ \frac{1}{2}[\hat{S},[\hat{S},\hat{H}_{\mathrm{rot}}]]
+ \cdots\, ,
\end{equation}
we choose $\hat{S}(t)$ to cancel the oscillating terms at first order:
\begin{equation}
- i \partial_t \hat{S}(t) + \hat{H}_{\mathrm{rot}}(t) = 0.
\end{equation}

This yields
\begin{equation}
\hat{S}(t) = \frac{1}{\Delta}
\left(
\hat{H}_{+} e^{i\Delta t} - \hat{H}_{-} e^{-i\Delta t}
\right).
\end{equation}

Substituting back, the first-order oscillating terms cancel, and the leading static contribution arises from
\begin{equation}
\hat{H}_{\mathrm{eff}} = \frac{1}{2}[\hat{S}, \hat{H}_{\mathrm{rot}}].
\end{equation}

Evaluating explicitly:
\begin{equation}
\hat{H}_{\mathrm{eff}} = \frac{1}{\Delta}[\hat{H}_{+}, \hat{H}_{-}]
+ \mathcal{O}\left(\frac{J^3}{\Delta^2}, \frac{\Omega_d^3}{\Delta^2}\right).
\end{equation}
We compute the commutator explicitly to find
\begin{equation}
[\hat{H}_{+}, \hat{H}_{-}]
=
[J \sigma_+^A \sigma_-^B + \tfrac{\Omega_d}{2}\sigma_+^A,\;
J \sigma_-^A \sigma_+^B + \tfrac{\Omega_d}{2}\sigma_-^A].
\end{equation}
Evaluating term by term:
\paragraph{$J^2$ term}
\begin{equation}
J^2 [\sigma_+^A \sigma_-^B, \sigma_-^A \sigma_+^B]
= J^2 \sigma_z^A \sigma_z^B.
\end{equation}

\paragraph{$J\Omega_d$ term}
\begin{equation}
J\Omega_d\, \sigma_z^A \sigma_x^B.
\end{equation}

\paragraph{$\Omega_d^2$ term}
\begin{equation}
\frac{\Omega_d^2}{4}[\sigma_+^A,\sigma_-^A]
= \frac{\Omega_d^2}{4}\sigma_z^A.
\end{equation}

Thus, the final form of the effective Hamiltonian is given by Eq.~\eqref{eq:Heff}:
\begin{equation}
\hat{H}_{\rm eff} =
\frac{J\Omega_d}{\Delta} \sigma_z^A \sigma_x^B
+ \frac{J^2}{\Delta} \sigma_z^A \sigma_z^B
+ \frac{\Omega_d^2}{4\Delta} I \otimes \sigma_x^B.
\end{equation}

The approximation $\hat{H}_{\rm eff}$ is controlled by the small parameters
\begin{equation}
\epsilon_J = \frac{J}{\Delta}, \quad
\epsilon_d = \frac{\Omega_d}{\Delta}.
\end{equation}
Neglected terms scale as
\begin{equation}
\mathcal{O}(\epsilon_J^3, \epsilon_d^3, \epsilon_J^2 \epsilon_d, \epsilon_J \epsilon_d^2),
\end{equation}
and are therefore small when $|J|, |\Omega_d| \ll |\Delta|$.

\subsection{Effect of weak anharmonicity}

For a weakly anharmonic transmon, the next excited state $\ket{2}$ is separated by $\alpha$ from the $\ket{1}\rightarrow\ket{2}$ transition. The drive can induce leakage with amplitude
$
\sim {\Omega_d}/(\Delta - \alpha)
$.
Similarly, coupling-induced processes scale as
$
\sim {J}/(\Delta - \alpha)
$.
Since typically $|\alpha| \ll |\Delta|$ but still $|\Delta - \alpha| \sim |\Delta|$, these processes remain perturbative and primarily renormalize the coefficients of the effective Hamiltonian without changing its operator structure at leading order.
Therefore, while higher levels quantitatively modify the interaction strengths, the effective CR interaction $\sigma_z^A \sigma_x^B$ arises already at the two-level approximation and remains the dominant entangling mechanism.
}

\begin{widetext}

\section{Perturbation theory}

\subsection{Matrix elements}

To estimate $\eta_{\alpha|\beta}$, perturbative analysis of the relevant matrix elements $\bra{0_T0_{\alpha}\beta}\hat{n}_{\alpha}\ket{1_T0_{\alpha}\beta}$ and $\bra{0_T0_{\alpha}\beta}\hat{n}_{T}\ket{1_T0_{\alpha}\beta}$ can be helpful. We will do so in the active two-qubit subspace, labeled as $\ket{i_{\alpha}j_T}$, and using shorthand notation $n_{ij}^T = \bra{i}\hat{n}_T\ket{j}, n_{ij}^\alpha = \bra{i}\hat{n}_\alpha\ket{j}$. Starting with the result from~\cite{nesterov2022cnot}, we use angular frequencies such that each $\omega_{ij}^{\alpha} = 2\pi f_{ij}^{\alpha}$. By excluding the zeroth order term in $J$, we get the first-order approximations
\begin{subequations}
    \begin{equation}
    \bra{00} \hat{n}_{\alpha} \ket{01} \approx -2\frac{J}{\hbar} n_{01}^T 
    \left[ 
    \frac{(n_{01}^{\alpha})^2 \, \omega_{01}^{\alpha}}{(\omega_{01}^{\alpha})^2 - (\omega_{01}^T)^2} 
    + \frac{(n_{03}^{\alpha})^2 \, \omega_{03}^{\alpha}}{(\omega_{03}^{\alpha})^2 - (\omega_{01}^T)^2} 
    \right],
    \end{equation}
    \begin{equation}
    \bra{00} \hat{n}_{T} \ket{01} \approx -2\frac{J}{\hbar} n_{01}^\alpha 
    \left[ 
    \frac{(n_{01}^{T})^2 \, \omega_{01}^{T}}{(\omega_{01}^{T})^2 - (\omega_{01}^\alpha)^2} 
    \right].
    \end{equation}
\end{subequations}
These expressions can be further broken down using $n_{01}^T \approx \sqrt[\leftroot{-2}\uproot{2}4]{E_J/32E_C}$ and $\omega_{01}^T \approx 2\pi(\sqrt{8E_CE_J}-E_C)$. With the same approach, additional matrix elements can be constructed as

\begin{subequations}
    \begin{equation}
    \bra{10} \hat{n}_{\alpha} \ket{11} \approx 2\frac{J}{\hbar} n_{01}^T 
    \left[ 
    \frac{(n_{01}^{\alpha})^2 \, \omega_{01}^{\alpha}}{(\omega_{01}^{\alpha})^2 - (\omega_{01}^T)^2} 
    + \frac{(n_{12}^{\alpha})^2 \, \omega_{12}^{\alpha}}{(\omega_{12}^{\alpha})^2 - (\omega_{01}^T)^2} 
    \right],
    \end{equation}
    \begin{equation}
    \bra{10} \hat{n}_{T} \ket{11} \approx 2\frac{J}{\hbar} n_{01}^{\alpha} 
    \left[ 
    \frac{(n_{01}^{T})^2 \, \omega_{01}^{T}}{(\omega_{01}^{T})^2 - (\omega_{01}^{\alpha})^2} 
    + \frac{(n_{12}^{T})^2 \, \omega_{12}^{T}}{(\omega_{12}^{T})^2 - (\omega_{01}^{\alpha})^2} 
    \right],
    \end{equation}
\end{subequations}
\begin{subequations}
    \begin{equation}
    \bra{00} \hat{n}_{\alpha} \ket{10} \approx 2\frac{J}{\hbar} n_{01}^{T} 
    \left[ 
    \frac{(n_{01}^{\alpha})^2 \, \omega_{01}^{\alpha}}{(\omega_{01}^{T})^2 - (\omega_{01}^{\alpha})^2} 
    + \frac{(n_{03}^{\alpha})^2 \, \omega_{03}^{\alpha}}{(\omega_{01}^{T})^2 - (\omega_{03}^{\alpha})^2} 
    \right],
    \end{equation}
    \begin{equation}
    \bra{00} \hat{n}_T \ket{10} \approx 2\frac{J}{\hbar} n_{01}^{\alpha} 
    \left[ 
    \frac{(n_{01}^{T})^2 \, \omega_{01}^{T}}{(\omega_{01}^{\alpha})^2 - (\omega_{01}^{T})^2} 
    \right],
    \end{equation}
\end{subequations}
\begin{subequations}
    \begin{equation}
    \bra{01} \hat{n}_{\alpha} \ket{11} \approx -2\frac{J}{\hbar} n_{01}^{T} 
    \left[ 
    \frac{(n_{01}^{\alpha})^2 \, \omega_{01}^{\alpha}}{(\omega_{01}^{T})^2 - (\omega_{01}^{\alpha})^2} 
    - \frac{(n_{12}^{\alpha})^2 \, \omega_{12}^{\alpha}}{(\omega_{01}^{T})^2 - (\omega_{12}^{\alpha})^2} 
    \right],
    \end{equation}
    \begin{equation}
    \bra{01} \hat{n}_T \ket{11} \approx -2\frac{J}{\hbar} n_{01}^{\alpha} 
    \left[ 
    \frac{(n_{01}^{T})^2 \, \omega_{01}^{T}}{(\omega_{01}^{\alpha})^2 - (\omega_{01}^{T})^2} 
    - \frac{(n_{12}^{T})^2 \, \omega_{12}^{T}}{(\omega_{01}^{\alpha})^2 - (\omega_{12}^{T})^2} 
    \right],
    \end{equation}
\end{subequations}
and further approximated using $n_{12}^{T} \approx \sqrt[\leftroot{-2}\uproot{2}4]{E_J/8E_C}$.

\subsection{Energy levels}
Dressed eigenenergies for our FTF system can be calculated either directly through numerical diagonalization or with perturbation theory. For the latter, second order perturbation theory is sufficient due to low qubit coupling strengths. For our case where $J = J_1 = J_2$ and $I = 0$, we only have pathways for matrix elements of $\hat{n}_1\hat{n}_T$ and $\hat{n}_2\hat{n}_T$. This leaves us with the simple expression
\begin{equation}
    E_{ijk}^{(2)} = J^2\sum_{p,q,l}{\left[\frac{\left| \bra{pql} \hat{n}_1\hat{n}_T \ket{ijk} \right|^2}{E_{ijk}^{(0)} - E_{pql}^{(0)}} + \frac{\left| \bra{pql} \hat{n}_2\hat{n}_T \ket{ijk} \right|^2}{E_{ijk}^{(0)} - E_{pql}^{(0)}}\right]},
\end{equation}
where $E^{(0)}$ denotes a bare energy state. Using the most dominant 4-12 terms per level to maintain precision, values of $E^{(2)}$ exhibit strong agreement with their numerically diagonalized counterparts. This accuracy is exemplified by perturbative estimates of the dressed transmon frequencies $\tilde{f_{01}^T}$ shown in Fig.~\ref{fig:fig5}.

\section{CX gates}
\label{appendix:cxphases}
Here, we analyze single qubit phase corrections required for a $CX_1$ gate, but the same idea can be applied symmetrically for a $CX_2$ gate. The necessary single-qubit rotations for correcting a $CX_1$ gate can be determined from the evolution operator
\begin{align*}
    \hat{U}_{\rm sim} = \begin{pmatrix}
    Xe^{ia} & x & x & x & x & x & x & x\\
    x & Xe^{ib} & x & x & x & x & x & x\\
    x & x & x & x & x & x & Xe^{ic} & x\\
    x & x & x & x & x & x & x & Xe^{id}\\
    x & x & x & x & Xe^{if} & x & x & x\\
    x & x & x & x & x & Xe^{ig} & x & x\\
    x & x & Xe^{ih} & x & x & x & x & x\\
    x & x & x & Xe^{ij} & x & x & x & x
    \end{pmatrix}.
\end{align*}
\end{widetext}
Ignoring the effects of small matrix elements labeled ``x'', $Z$-rotations of the control and spectator qubits commute with $U_{\rm sim}$. Thus, the two fluxoniums each require a single rotation, while the transmon benefits from having both pre-CR and post-CR $R_Z$ gates. The sequence for a corrected gate is then $\hat{U}_{\rm rot} = (R_Z(\theta_2) \otimes R_Z(\theta_3) \otimes R_Z(\theta_4))\hat{U}_{\rm sim}(R_Z(\theta_1) \otimes I \otimes I)$, using standard single-qubit $Z$-rotations given by
\begin{align}
    R_Z(\theta) = \begin{pmatrix}
    e^{-i\theta/2} & 0\\
    0 & e^{i\theta/2} 
    \end{pmatrix}.
\end{align}
In a scaled grid of alternating fluxoniums and transmons, spectators requiring phase correction may raise concerns about the feasibility of simultaneous gates on FT pairs with shared spectators. To address this, we remind the reader that $R_Z$ gates both commute with each other and can be effectively tracked classically. With this in mind, the only overhead of a $CX_{\alpha}$ gate in the experimental case requires learning of the accumulated phase on each involved qubit. Angles $\theta_i$ can be chosen by numerical fidelity optimization, as is used to generate our data throughout this paper, or with a simplified analytical approach. For the latter, we consider phase correction terms corresponding to matrix elements when F2 is in state $\ket0$ ($\ket1$). To minimize spectator qubit bias, we then average the necessary rotation angles in these two subspaces. Explicitly, these angles correspond to
\begin{subequations}
\begin{align}
\theta_{1} = \frac{1}{4} (-f - c + h + a) + \frac{1}{4} (-g - d + j + b) + \pi,\\
\theta_{2} = \frac{1}{4} (-f - h + c + a) + \frac{1}{4} (-g - j + d + b) - \pi,\\
\theta_{3} = \frac{1}{4} (-h - c + a + f) + \frac{1}{4} (-d - j + g + b) - \pi.
\end{align}
\end{subequations}
The angle for F2 is trivially $\theta_4 = (a - b)$. This approach can be used alongside $CX_\alpha$ gate tomography data on a physical device to construct the necessary virtual rotations~\cite{McKay2017VZ}.

\section{Error budget}\label{appendix:errorbudget}

\subsection{2-qubit system}
To construct a \textsc{cnot} gate error budget, we define $\mathcal{E}_{bright}$, $\mathcal{E}_{\rm dark},$ and $\mathcal{E}_{\rm leak}$. These error categorizations come from a decomposition of $\mathcal{E} = 1 - \mathcal{F}_{\rm coh}$ where $\mathcal{F}_{\rm coh}$ is calculated with Eq.~\eqref{eq:fidelity}. Explicitly, this begins with
\begin{equation}
    \mathcal{E} = 1 - \frac{\mathrm{Tr}(\hat{U}^\dagger \hat{U})}{20} - \frac{\left| \mathrm{Tr}(\hat{U}^\dagger_{id} \hat{U}) \right|^2}{20}.
\end{equation}
Leakage error is separable from the rest as
\begin{equation}
    \mathcal{E}_{\rm leak} = \frac{1}{5} - \frac{{\mathrm{Tr} }\{\hat{U}^{\dagger} \hat{U}\}}{20},
\end{equation}
which leaves
\begin{equation}
    \mathcal{E}_{comp} = \frac{4}{5} - \frac{\left| \mathrm{Tr}(\hat{U}^\dagger_{id} \hat{U}) \right|^2}{20}.
    \label{eq:2qcomperror}
\end{equation}
For categorization, we split $\mathrm{Tr}(\hat{U}_{id}^{\dag}\hat{U})$ into $D + B$, where $D$ ($B$) accounts for the matrix elements of two correct transitions, which are ideally dark (bright). Since $U_{id}$ only has matrix elements of $0$ and $1$, D and B account for all significant terms in the low-error limit. Explicitly, these terms are
\begin{subequations}
    \begin{equation}
        D = \Pi_{00 \to 00} + \Pi_{01 \to 01},
    \end{equation}
    \begin{equation}
        B = \Pi_{10 \to 11} + \Pi_{11 \to 10}.
    \end{equation}
\end{subequations}
Although matrix elements $\Pi_s$ are complex, $ZZ$ is constant in the two-qubit case, and we can eliminate its effect with local phase rotations~\cite{NesterovTwoPhoton2021}. As such, $\Pi_s = \left|\Pi_s\right|$ can be set by convention. By substituting D and B into Eq.~\eqref{eq:2qcomperror}, we have
\begin{equation}
    \mathcal{E}_{\rm comp} = \frac{4}{5} - \frac{\left|D + B\right|^2}{20}.
\end{equation}
Replacing $D$ and $B$ by their corresponding error terms $\mathcal{E}_D = 2 - D$ and $\mathcal{E}_B = 2 - B$ gives
\begin{equation}
    \mathcal{E}_{\rm comp} = \frac{4}{5} - \frac{(2 - \mathcal{E}_D)^2 + 2(2-\mathcal{E}_D)(2-\mathcal{E}_B) + (2-\mathcal{E_B})^2}{20}.
\end{equation}
Since we are mostly concerned with convergent accuracy in the low error limit, we drop second-order terms to obtain
\begin{equation}
    \mathcal{E}_{\rm comp} = \frac{2 \mathcal{E}_D + 2\mathcal{E}_B}{5}.
\end{equation}
Finally, by defining $\mathcal{E}_{\rm dark} = \frac{2}{5}\mathcal{E}_B$ and $\mathcal{E}_{\rm flip} = \frac{2}{5}\mathcal{E}_B$, we recover the error terms used in Eq.~\eqref{eq:2qerrorterms}.

Analysis of error contributions can also be done by grouping similar error transitions. In particular, following~\cite{nesterov2022cnot}, we can construct terms such that
\begin{equation}
\label{eq:2qerrcat}
\mathcal{E} = \mathcal{E}_{\mathrm{ctrl},1} + \mathcal{E}_{\mathrm{ctrl},2}+ \mathcal{E}_{\mathrm{dark}} +\mathcal{E}_{\mathrm{bright}}+\mathcal{E}_{\rm leak}\,. 
\end{equation}
Individual terms are evaluated using matrix elements of the evolution operator using the notation
\begin{equation}
P_{ab\to a'b'} = |\bra{a'b'}\hat{U}_{\rm sim}\ket{ab}|^2.
\end{equation}
We group transitions based on their dynamics. The first two error terms are associated with a flip of the control qubit. In particular, $\mathcal{E}_{\mathrm{ctrl,1}}$ corresponds to only a control qubit flip with correct transmon behavior, while $\mathcal{E}_{\mathrm{ctrl,2}}$ corresponds to incorrect flipping behavior of both qubits. Explicitly, they are defined as
\begin{subequations}
    \begin{equation}
    \mathcal{E}_{\mathrm{ctrl,1}} = (P_{00\to 10}+P_{10\to 00} + P_{01\to 10}+P_{10\to 01})/5,
    \end{equation} 
    \begin{equation}
    \mathcal{E}_{\mathrm{ctrl,2 }}= (P_{01\to 11}+P_{11\to 01} + P_{00\to 11}+P_{11\to 00})/5.
    \end{equation}
\end{subequations}
$\mathcal{E}_{\rm dark}$ quantifies the rate of target qubit flips from a dark initial state as
\begin{equation}
\mathcal{E}_{\rm dark} =(P_{00\to 01}+P_{01\to 00})/5.
\end{equation}
In contrast, $\mathcal{E}_{\rm bright}$ quantifies the rate of target qubit non-flips from a bright initial state as
\begin{equation}
\mathcal{E}_{\rm bright} = (P_{10\to 10}+P_{11\to 11})/5.
\end{equation} 
Finally, in addition to the errors within the computational subspace, we evaluate the leakage error as 
\begin{equation}
\mathcal{E}_{\rm leak} = 1-{\mathrm{Tr} }\{\hat{U}_{\rm sim}^{\dagger} \hat{U}_{\rm sim}\}/4.
\end{equation}
Due to their similar dynamics, individual transitions within $\mathcal{E}_{\mathrm{ctrl},1}$ and $\mathcal{E}_{\mathrm{ctrl},2}$ tend to have similar numerical values. As a result, each value provides fairly detailed information on the magnitudes of its corresponding matrix elements. In the type of system we consider, however, large detuning between control and target qubits prevents these terms from ever dominating. On the other hand, $\mathcal{E}_{\rm bright}$ and $\mathcal{E}_{\rm dark}$ capture the main error mechanism in our two-qubit system based on a given initial state. Along with the leakage term, the idea of grouping errors based on having dark or bright initial states is retained in our new error budget.

\subsection{3-qubit system}

We derive error terms $\mathcal{E}_{\rm dark}$ and $\mathcal{E}_{\rm flip}$ for the three-qubit case, extending our novel method from the two-qubit case. For simplicity, we will now refer to the evolution operators $CX_{\alpha}$ as $\hat{U}$. Starting with $\mathcal{E} = 1 - \mathcal{F}_{\rm coh}$, we write
\begin{equation}
    \mathcal{E} = 1 - \frac{\mathrm{Tr}(\hat{U}^{\dag}\hat{U})}{72} - \frac{\left|\mathrm{Tr}(\hat{U}_{id}^{\dag}\hat{U})\right|^2}{72}.
\end{equation}
Substituting in the leakage error 
\begin{equation}
    \mathcal{E}_{\rm leak} = \frac{1}{9} - \frac{\mathrm{Tr}\{\hat{U}^{\dagger} \hat{U}\}}{72},
\end{equation}
 yields
\begin{equation}
    \mathcal{E} = \mathcal{E}_{\rm leak} + \frac{8}{9} - \frac{\left|\mathrm{Tr}(\hat{U}_{id}^{\dag}\hat{U})\right|^2}{72},
\end{equation}
separating errors in the computational subspace from leakage terms.
In the computational subspace, we define matrix elements $D$ and $B$ as sums of transition amplitudes $\Pi_s$
\begin{subequations}
    \begin{equation}
        D = \sum_{i,j=0}^1 \Pi_{i0j \to i0j},
    \end{equation}
    \begin{equation}
        B = \sum_{i,j=0}^1 \Pi_{i1j \to \bar{i}1j},
    \end{equation}
\end{subequations}
where now each matrix element is strictly complex and defined as $\Pi_s = \exp(i \phi_s)|\Pi_s|$. Unlike the two-qubit case, $ZZ$ coupling between the two active qubits is dependent on the spectator qubit state and therefore not static. Consequently, we cannot fully eliminate it with local phase rotations, and the resulting phase mismatch errors must be considered. For consistency with the two-qubit approach, we will separate such errors into an additional ``phase'' term. Further, computational error can be represented as
\begin{equation}
    \mathcal{E}_{\rm comp} = \frac{8}{9} - \frac{|D + B|^2}{72},
\end{equation}
with $\mathcal{E}_D = 4 - D$, and $\mathcal{E}_B = 4 - B$.
Expanding and keeping only first-order error terms, we get
\begin{equation}
    |D + B|^2 \approx 64 - 8(\mathcal{E}_D + \mathcal{E}_D^* + \mathcal{E}_B + \mathcal{E}_B^*).
\end{equation}
Thus, the computational error contribution becomes
\begin{equation}
    \mathcal{E}_{\rm comp} = \frac{(\mathcal{E}_D + \mathcal{E}_D^* + \mathcal{E}_B + \mathcal{E}_B^*)}{9}.
\end{equation}
Since $\mathcal{E}_D + \mathcal{E}_D^* = 2 \mathrm{Re}(\mathcal{E}_D)$ and $\mathcal{E}_B + \mathcal{E}_B^* = 2 \mathrm{Re}(\mathcal{E}_B)$, we have
\begin{equation}
    \mathcal{E}_{\rm comp} = \frac{2 \mathrm{Re}(\mathcal{E}_D) + 2 \mathrm{Re}(\mathcal{E}_B)}{9}.
\end{equation}
To first obtain the dark and flip terms without considering phase-type errors, we take $\mathrm{Re}(\Pi_s) = \left|\Pi_s\right|$. Then, separating $\mathcal{E}_{\rm comp}$ as before, we recover the final error budget terms
\begin{subequations}
    \begin{equation}
        \mathcal{E}_{\rm dark} = \frac{2}{9} \sum_{i,j=0}^1 (1 - \left|\Pi_{i0j \to i0j}\right|),
    \end{equation}
    \begin{equation}
        \mathcal{E}_{\rm flip} = \frac{2}{9} \sum_{i,j=0}^1 (1 - \left|\Pi_{i1j \to \bar{i}1j}\right|).
    \end{equation}
\end{subequations}
To now correct for phase $\phi_s$ in the major matrix elements $\mathrm{Re}(\Pi_s) = \left|\Pi_s\right|\cos(\phi_s)$, we also define
\begin{equation}
    \mathcal{E}_{\rm phase} = \mathcal{E}_{\rm d,phase} +
    \mathcal{E}_{\rm b,phase} +
    \mathcal{E}_{\rm imag}.
\end{equation}
Here, $\mathcal{E}_{\rm d,phase}$ and $\mathcal{E}_{\rm b,phase}$ are phase-dependent corrections to the dark and bright terms, respectively. The term $\mathcal{E}_{\rm imag}$ arises from the presence of imaginary components $\mathrm{Im}(\Pi_s) = \left|\Pi_s\right|i\sin(\phi_s)$ in each matrix element. These contributions are explicitly defined as
\begin{subequations}
    \begin{equation}
        \begin{split}
        \mathcal{E}_{\rm d,phase} = \frac{2}{9} \sum_{i,j=0}^1{\left|\Pi_{i0j \to i0j}\right|(1 - \cos{\phi_{i0j}})},
        \end{split}
    \end{equation}
    \begin{equation}
        \begin{split}
        \mathcal{E}_{\rm b,phase} = \frac{2}{9} \sum_{i,j=0}^1{\left|\Pi_{i1j \to \bar{i}1j}\right|(1 - \cos{\phi_{i1j}})},
        \end{split}
    \end{equation}
    \begin{equation}
        \begin{split}
        \mathcal{E}_{\rm imag} = - &\frac{1}{72}\bigg[\sum_{i,j=0}^1\big[ \left|\Pi_{i0j \to i0j}\right| \sin{\phi_{i0j}}
        \\
        +\; &\left|\Pi_{i1j \to \bar{i}1j}\right| \sin{\phi_{i1j}}\big]\bigg]^2.
        \end{split}
    \end{equation}
\end{subequations}
We now have the complete 3-qubit error budget found in Eq.~\eqref{eq:darkflipterms}.

\begin{widetext}

\section{Parity checks}

Our ideal parity check without intermediate phase corrections is represented by the evolution operator
\begin{align*}
    \hat{U}_{\rm PC} = \begin{pmatrix}
    Xe^{ia} & x & x & x & x & x & x & x\\
    x & x & x & x & x & Xe^{ib} & x & x\\
    x & x & x & x & x & x & Xe^{ic} & x\\
    x & x & x & Xe^{id} & x & x & x & x\\
    x & x & x & x & Xe^{if} & x & x & x\\
    x & Xe^{ig} & x & x & x & x & x & x\\
    x & x & Xe^{ih} & x & x & x & x & x\\
    x & x & x & x & x & x & x & Xe^{ij}
\end{pmatrix}.
\end{align*}
\end{widetext}
To avoid several unnecessary phase-correction rotations, $CX_1$ and $CX_2$ pulses can be applied sequentially before a single set of $R_Z$ gates restores important phase information. Since T is measured in the $Z$-basis, its phase information following a parity check can be ignored. As such, we only correct the phase in the subsystem involving F1 and F2. Given preservation of unitarity and the transmon starting in $\ket0$, the effective gate on the fluxoniums is
\begin{align}
\hat{U}_{\rm FF} = \begin{pmatrix}
Xe^{ia} & x & x & x\\
x & Xe^{ig} & x & x\\
x & x & Xe^{ih} & x\\
x & x & x & Xe^{id}
\end{pmatrix}.
\end{align}
We choose $\theta_1 = a/2 - d/2 + g/2 - h/2$ and $\theta_2 = a/2 - d/2 - g/2 + h/2$, as angles for the sequence $\hat{U}_{FF}^{\rm (rot)} =(R_Z(\theta_1)\otimes R_Z(\theta_2))\hat{U}_{FF}$, such that
\begin{align}
    \hat{U}_{FF}^{\rm (rot)} = \begin{pmatrix}
    Xe^{i\frac{(a+d)}{2}} & x & x & x\\
    x & Xe^{i\frac{(g+h)}{2}} & x & x\\
    x & x & Xe^{i\frac{(g+h)}{2}} & x\\
    x & x & x & Xe^{i\frac{(a+d)}{2}}
    \end{pmatrix}.
\end{align}
Now, the even and odd fluxonium parity subspaces have global phase accumulation factors of $(a+d)/2$ and $(g+h)/2$, respectively. By post-selecting based on measurement of the transmon, $\hat{U}_{FF}^{\rm (rot)}$ can be reduced to a fully coherent $2\times2$ matrix for the subspace of correct parity.
\section{Stability with respect to Josephson energies}
\label{sec:EJStability}
\begin{table}[h]
\centering
\begin{tabular}{||c c c | c ||}%
 \hline
 $E_{J,1}/h$ (GHz) & $E_{J,2}/h$ (GHz) & $E_{J,T}/h$ (GHz) & $CX_1$ Error ($10^{-5}$)\\ [0.5ex] 
 \hline\hline
 4.21 & 4.2 & 18 & 7.67 \\
 \hline
 4.21 & 4 & 18 & 6.12 \\
 \hline
 4.21 & 3.8 & 18 & 5.11 \\
 \hline
 4.01 & 4 & 18 & 6.38 \\
 \hline
 4.01 & 3.8 & 18 & 5.01 \\
 \hline
 3.81 & 3.8 & 18 & 5.85 \\
 \hline
 4.21 & 4.2 & 17 & 8.50 \\
 \hline
 4.21 & 4 & 17 & 6.71 \\
 \hline
 4.21 & 3.8 & 17 & 5.56 \\
 \hline
 4.01 & 4 & 17 & 7.15 \\
 \hline
 4.01 & 3.8 & 17 & 5.66 \\
 \hline
 3.81 & 3.8 & 17 & 5.77 \\
 \hline
4.21 & 4.2 & 16 & 30.6 \\
 \hline
 4.21 & 4 & 16 & 29.6 \\
 \hline
 4.21 & 3.8 & 16 & 27.1 \\
 \hline
 4.01 & 4 & 16 & 10.5 \\
 \hline
 4.01 & 3.8 & 16 & 8.70 \\
 \hline
 3.81 & 3.8 & 16 & 9.45 \\
 \hline
\end{tabular}
\caption{Total $CX_1$ error ($10^{-5}$) for varied qubit Josephson energies (GHz) in an optimized 50 ns gate.}
\label{tab:tab2}
\end{table}
With current state-of-the-art fabrication, variance in targeted qubit $E_J$ values is non-negligible~\cite{hertzberg2021laser, osman2021simplified, Wang2022}. For F1 and F2, we find that small percent variances in Josephson energy do not give rise to any resonances that spoil gate performance. In contrast, the transmon Josephson energy is much larger and thus even small relative variations can shift its operating frequency into resonance with a fluxonium leakage transition. Increasing $E_{J,T}/h$ above our studied value of $18$ GHz leads to a regime where optimal gates are leakage-dominated regardless of total pulse time. This is a result of the active fluxonium $\ket0 \xrightarrow{} \ket4$ transition coming into resonance with the targeted bright transmon transition.
\begin{figure}[ht]
\includegraphics[width=0.375\textwidth]{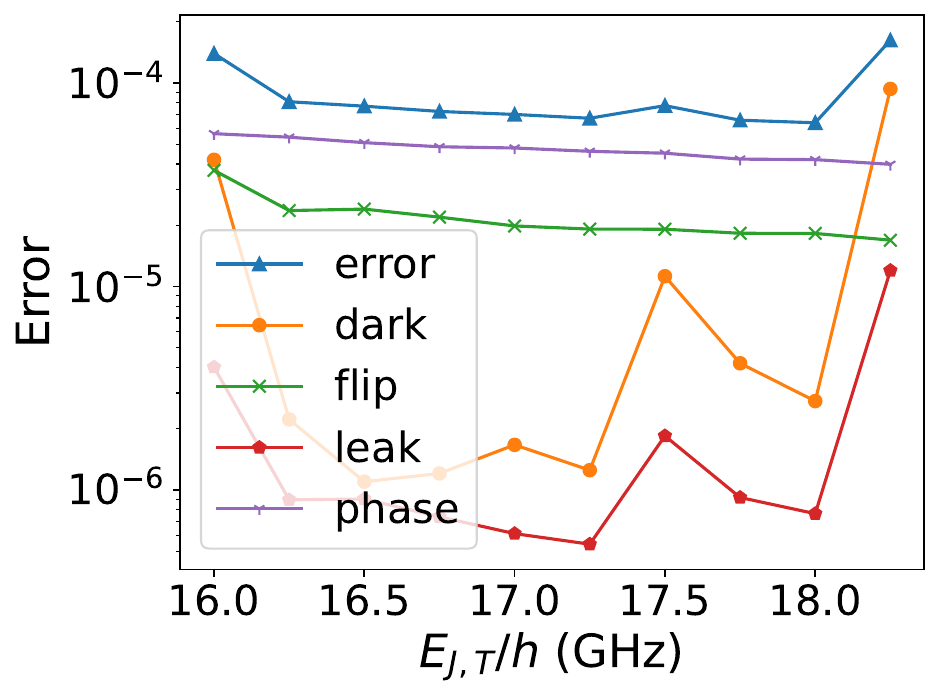}
\caption{50 ns $CX_1$ gate error as a function of transmon Josephson energy with all other system parameters taken from Table~\ref{tab:tab1}.}\label{fig:fig12}
\end{figure}
Due to our coupling of fluxonium-transmon pairs, emergent resonances at larger $E_{J,T}$ prohibit high-fidelity gates that avoid driving leakage. To experimentally circumvent this, we suggest designing the transmon such that $18$ GHz would be the upper limit for $E_{J,T}/h$ within fabrication tolerances. Going much lower than this can lead to other unwanted resonances, higher transmon charge sensitivity, and less detuning between the $\ket0 \leftrightarrow{} \ket1$ transitions of the control and target qubits. For our purposes, we assume $E_{J,1}/h = 4, E_{J,2}/h = 4, E_{J,T}/h = 17$ to be targeted experimental values for which a variance of less than $5\%$ can be expected. Choosing resulting high, on target, and low benchmark values, we optimize $CX_1$ drive parameters and report the minimal possible gate error in each system. Since F1 and F2 are identical aside from their Josephson Energies, we only list distinct combinations of $E_{J,1}$ and $E_{J,2}$. To avoid degeneracy in numerical diagonalization when fluxonium Josephson energies are identical, we also introduce a $0.01$ GHz bias to F1, which has a negligible effect on any dynamics. Within our proposed window of $E_J$ variation, all combinations demonstrate fairly high optimal-parameter \textsc{cnot} fidelity.
\bibliography{main}
\end{document}